\documentstyle[psfig]{mn}

\newif\ifAMStwofonts

\newcommand{\be}{\begin{equation}}
\newcommand{\ee}{\end{equation}}
\newcommand{\bea}{\begin{eqnarray}}
\newcommand{\eea}{\end{eqnarray}}
\newcommand{\x}{${\bf x}$}
\newcommand{\mx}{{\bf x}}

\newcommand{\q}{${\bf q}$}
\newcommand{\mq}{{\bf q}}
\renewcommand{\S}{${\bf S}$}
\newcommand{\mS}{{\bf S}}
\newcommand{\s}{${\bf s}$}
\newcommand{\ms}{{\bf s}}
\newcommand{\mpc}{$h^{-1}\; {\rm Mpc}$}

\ifoldfss
  \ifCUPmtlplainloaded \else
    \NewTextAlphabet{textbfit} {cmbxti10} {}
    \NewTextAlphabet{textbfss} {cmssbx10} {}
    \NewMathAlphabet{mathbfit} {cmbxti10} {} 
    \NewMathAlphabet{mathbfss} {cmssbx10} {} 
  \fi
  \ifAMStwofonts
    \ifCUPmtlplainloaded \else
      \NewSymbolFont{upmath} {eurm10}
      \NewSymbolFont{AMSa} {msam10}
      \NewMathSymbol{\upi}     {0}{upmath}{19}
      \NewMathSymbol{\umu}     {0}{upmath}{16}
      \NewMathSymbol{\upartial}{0}{upmath}{40}
      \NewMathSymbol{\leqslant}{3}{AMSa}{36}
      \NewMathSymbol{\geqslant}{3}{AMSa}{3E}

       \let\le=\leqslant
       \let\ge=\geqslant
    \fi
  \fi
\fi 

\ifnfssone
  \newmathalphabet{\mathit}
  \addtoversion{normal}{\mathit}{cmr}{m}{it}
  \addtoversion{bold}{\mathit}{cmr}{bx}{it}
  \newmathalphabet{\mathbfit} 
  \addtoversion{normal}{\mathbfit}{cmr}{bx}{it}
  \addtoversion{bold}{\mathbfit}{cmr}{bx}{it}
  \newmathalphabet{\mathbfss} 
  \addtoversion{normal}{\mathbfss}{cmss}{bx}{n}
  \addtoversion{bold}{\mathbfss}{cmss}{bx}{n}
  \ifAMStwofonts
    \ifCUPmtlplainloaded \else
      \UseAMStwoboldmath
      \makeatletter
      \new@mathgroup\upmath@group
      \define@mathgroup\mv@normal\upmath@group{eur}{m}{n}
      \define@mathgroup\mv@bold\upmath@group{eur}{b}{n}
      \edef\UPM{\hexnumber\upmath@group}
      \new@mathgroup\amsa@group
      \define@mathgroup\mv@normal\amsa@group{msa}{m}{n}
      \define@mathgroup\mv@bold\amsa@group{msa}{m}{n}
      \edef\AMSa{\hexnumber\amsa@group}
      \makeatother
      \mathchardef\upi="0\UPM19
      \mathchardef\umu="0\UPM16
      \mathchardef\upartial="0\UPM40
      \mathchardef\leqslant="3\AMSa36
      \mathchardef\geqslant="3\AMSa3E

       \let\le=\leqslant
       \let\ge=\geqslant
    \fi
  \fi
\fi 

\ifnfsstwo
  \DeclareMathAlphabet{\mathbfit}{OT1}{cmr}{bx}{it}
  \SetMathAlphabet\mathbfit{bold}{OT1}{cmr}{bx}{it}
  \DeclareMathAlphabet{\mathbfss}{OT1}{cmss}{bx}{n}
  \SetMathAlphabet\mathbfss{bold}{OT1}{cmss}{bx}{n}
  \ifAMStwofonts
    \ifCUPmtlplainloaded \else
      \DeclareSymbolFont{UPM}{U}{eur}{m}{n}
      \SetSymbolFont{UPM}{bold}{U}{eur}{b}{n}
      \DeclareSymbolFont{AMSa}{U}{msa}{m}{n}
      \DeclareMathSymbol{\upi}{0}{UPM}{"19}
      \DeclareMathSymbol{\umu}{0}{UPM}{"16}
      \DeclareMathSymbol{\upartial}{0}{UPM}{"40}
      \DeclareMathSymbol{\leqslant}{3}{AMSa}{"36}
      \DeclareMathSymbol{\geqslant}{3}{AMSa}{"3E}

       \let\le=\leqslant
       \let\ge=\geqslant
    \fi
  \fi
\fi 

\ifCUPmtlplainloaded \else
  \ifAMStwofonts \else 
    \def\upi{\pi}
    \def\umu{\mu}
    \def\upartial{\partial}
  \fi
\fi

\title[PDF of Initial Conditions from the PSCz catalogue]{The 1-point
PDF of the Initial Conditions of our Local Universe from the IRAS PSC
redshift catalogue}

\author[P. Monaco et al.]
       {P. Monaco$^{1,2}$, G. Efstathiou$^2$, S. J. Maddox$^2$, 
	E. Branchini$^3$, \cr C. S. Frenk$^4$, R. G. McMahon$^2$, 
        S. J. Oliver$^5$, M. Rowan-Robinson$^5$, \cr W. Saunders$^6$, 
	W. J. Sutherland$^7$, H. Tadros$^7$, S. D. M. White$^8$\\ 
        $^1$Dipartimento di Astronomia,	Universit\`a di Trieste, via Tiepolo 11, 34131 Trieste, Italy\\ 
	$^2$Institute of Astronomy, Madingley Road, Cambridge CB3 0HA, UK\\ 
	$^3$Kapteyn Sterrewacht, Rijksuniversiteit Groningen, Postbus 800, 9700, AV Groningen, The Netherlands\\
	$^4$Department of Physics, University of Durham, DH1 3LE, UK\\
	$^5$Blackett Laboratory, Imperial College, Prince Consort Road, London SW7 2BZ, UK\\
	$^6$Institute for Astronomy, Blackford Hill, Edinburgh EH9 3RJ, UK \\ 
	$^7$Nuclear and Astrophysics Laboratory, Keble Road, Oxford OX1 3RH, UK \\
	$^8$ MPI-Astrophysik,Karl-Schwarzschild-Strasse 1, Garching bei Munchen, Germany D-85740
}

\date{Received 2000}

\pagerange{\pageref{firstpage}--\pageref{lastpage}}
\pubyear{2000}

\begin{document}

\maketitle

\label{firstpage}

\begin{abstract}

The algorithm ZTRACE of Monaco \& Efstathiou (1999) is applied to the
IRAS PSCz catalogue to reconstruct the initial conditions of our local
Universe with a resolution down to $\sim$5 \mpc.  The 1-point PDF of
the reconstructed initial conditions is consistent with the
assumptions that (i) IRAS galaxies trace mass on scales of $\sim$5
\mpc, and (ii) the statistics of primordial density fluctuations is
Gaussian.  We use simulated PSCz catalogues, constructed from N-body
simulations with Gaussian initial conditions, to show that local
non-linear bias can cause the recovered initial PDF (assuming no bias)
to be non-Gaussian.  However, for plausible bias models, the
distortions of the recovered PDF would be difficult to detect using
the volume finely sampled by the PSCz catalogue.  So, for Gaussian
initial conditions, a range of bias models remain compatible with our
PSCz reconstruction results.

\end{abstract}


\section{Introduction}

In the gravitational instability scenario, the large-scale structure
of the Universe is formed by the growth of small fluctuations
superimposed on a homogeneous background.  In most inflationary
models, these fluctuations are usually predicted to follow Gaussian
statistics (see, e.g., Linde, 1990).  Non-Gaussian initial conditions
are however predicted by models based on topological defects (see,
e.g., Brandenberger 1998), as well as by some inflationary models
(see, e.g., Salopek 1999).  At the present time, perturbations at
galactic or smaller scales are highly non-linear, while the
large-scale distribution of galaxies is still in the linear or mildly
non-linear regime.  For these fluctuations, the mapping of the mass
distribution from the final to the initial configuration is
single-valued.  The gravitational evolution of the matter field can be
inverted, even though at relatively small scales the evolution is
beyond the linear stage and strong non-Gaussian features in the
density field are already present.

Under the assumption that galaxies trace the large-scale distribution
of mass in a specified way, it is possible to recover the initial
perturbation field which has given rise to the observed galaxy
distribution.  It is interesting to try to recover the initial
conditions of our local Universe, not simply for understanding our
local cosmography, but for other reasons as well.  The inversion
itself does not require any assumption concerning the detailed
statistical properties of the density field.  A reconstruction of
initial conditions can therefore be used to test the assumption of
Gaussianity.  Furthermore, it can be used in N-body simulations to
mimic the growth of clustering in our local Universe.  This type of
simulations have many uses, e.g. to produce predictions of our local
peculiar velocity field, to quantify Malmquist biases in peculiar
velocity measurements etc. (see, e.g., Kolatt et al, 1996; Narayanan
et al. 1999).

The reconstruction of initial conditions is hampered by the well-known
facts that galaxies are biased tracers of the matter density, and that
the details of this bias are poorly known (see, e.g., Tegmark \&
Peebles 1998; Catelan et al. 1998; Lahav \& Dekel 1999; Pearce et
al. 1999; Benson et al. 1999; Somerville et al. 1999; Seljak 2000;
Sigad, Branchini \& Dekel 2000).  A model-independent inclusion of
bias in a reconstruction algorithm is not possible at present.
Besides, for initial Gaussian perturbations the non-Gaussian
statistics of the matter density field is determined by gravitational
evolution.  Galaxy bias influences the statistics of the galaxy
density field.  A reconstruction procedure applied to a biased galaxy
density field but performed assuming no bias can correct only for
gravitationally induced non-Gaussianities, thus letting bias induce
non-Gaussian features in the recovered initial conditions.  In other
words, there is a degeneracy between initial non-Gaussianity and the
effects of galaxy bias.  The Gaussianity of initial conditions is much
better tested on very large scales by the cosmic microwave background
(CMB).  A careful analysis of the COBE data has lead to a detection of
small non-Gaussianity (Ferreira, Magueijo \& Gorski 1998), which is
likely to be an instrumental effect (Bromley \& Tegmark 1999; Banday,
Zaroubi \& Gorski 1999; Contaldi et al. 1999).  Future measurements of
Boomerang, MAP and PLANCK will either detect non-Gaussianity or
strengthen the limits on it.  Such possible non-Gaussian features are
anyway small.  With plausible extrapolation, they are likely to remain
considerably smaller, at the $\sim$10 \mpc\ scale, than the signal
that galaxy bias can induce in a reconstruction scheme as the one
presented here (see Verde et al. 1999).  As a consequence, the
statistics of the reconstructed initial conditions can be used to give
new valuable constraints on galaxy bias models, once the Gaussianity
of initial conditions is either assumed or tightly constrained by the
future CMB measurements.

The mildly non-linear evolution of the matter field is well modeled by
the Zel'dovich (1970) approximation.  It describes the evolution of
the growing mode of perturbations into the non-linear regime, as long
as the flow is still laminar.  This restriction to the growing-mode
dynamics gives an advantage over a backward N-body simulation, in
which noise is amplified as a decaying mode.  An inversion of the
Zel'dovich mapping was proposed by Nusser \& Dekel (1992).  Their
`time machine' code requires the velocity potential field in
real-space as an input.  When applied to a galaxy catalogue, the
initial conditions are not self-consistently obtained from the
observed quantity, i.e. the galaxy density field in redshift space.
Other methods for reconstructing the initial conditions, related to
that of Nusser \& Dekel (1992), were proposed by Gramann (1993) and
Taylor \& Rowan-Robinson (1993).  Croft \& Gatza\~naga (1997)
developed PIZA, a method based on the least-action principle (Peebles
1989) and the Zel'dovich approximation.  The same least-action
principle has been exploited to construct efficient reconstruction
algorithms by Nusser \& Branchini (1999) and by Golberg and Spergel
(1999).  Weinberg (1992) and Narayanan \& Weinberg (1998) imposed
Gaussianity of the recovered initial conditions to improve the
performance of a reconstruction algorithm applied to evolved Gaussian
fields.  Narayanan \& Croft (1999) compared the performance of
different reconstruction algorithms to N-body simulations. According
to them, the most accurate reconstruction schemes are that of
Narayanan \& Weinberg (1998) and PIZA.

The correctness of the reconstruction of the initial conditions relies
on the accuracy of the input galaxy sample, and on the extent of the
sampled volume.  In this regard, the uniformity of the galaxy
selection and the sky coverage are critical aspects.  For this reason,
the galaxy samples used to construct the galaxy density field have
been selected from the objects contained in the all-sky point source
catalogue (PSC) of the IRAS satellite, because they are observed on
the whole sky by the same instrument, and IR-selected samples are less
affected by dust extinction due to the Galactic Plane (although good
all-sky optical catalogues as the ORS (Santiago et al. 1995) or the
NOG (Marinoni et al. 1999) are now available).  The far-IR selection
is biased against spheroidal objects, so that IRAS galaxies tend to
miss rich structures (even though at scales larger than $\sim 5$ \mpc\
this effect is small, see Baker et al. 1999).  The QDOT redshift
catalogue (Lawrence et al. 1999) was defined by sparsely sampling the
PSC galactic objects with 60$\mu$ flux larger than 0.6 Jy (the
completeness limit).  Fisher et al. (1995) defined their sample by
limiting the PSC to a 60$\mu$ flux of 1.2 Jy.  More recently, a
redshift catalogue of all the PSC galaxies (hereafter PSCz; Saunders
et al. 2000) has been completed.  This catalogue provides the best
presently available sample to quantify the galaxy density field of our
local Universe.

To reconstruct the initial conditions of our local Universe, Nusser,
Dekel \& Yahil (1995) used the real-space density field of Yahil et
al. (1991), obtained from the IRAS 1.2 Jy redshift catalogue, smoothed
with a Gaussian filter of width 10 \mpc, using an iterative technique
based on linear theory.  The peculiar velocity potential, obtained
from the real-space density field by means of an average non-linear
velocity-density relation, was given as an input to the Nusser \&
Dekel's (1992) time machine.  The 1-point probability distribution
function (hereafter PDF) of their recovered initial conditions was
consistent with a Gaussian.  A reconstruction of the initial
conditions was performed also by Kolatt et al. (1996), using the 1.2
Jy catalogue, and Narayanan et al. (1999), using the PSCz catalogue.
In both papers the Gaussianity of the initial conditions was not found
from data but forced, following the Gaussianization technique of
Weinberg (1992) or Narayanan \& Weinberg (1999).  The recovered
initial conditions were used to run N-body simulations to construct
mock peculiar velocity catalogues (Kolatt et al. 1996) or to study the
effect of biased galaxy formation in the reconstruction (Narayanan et
al. 1999).

Monaco \& Efstathiou (1999, hereafter ME99) have recently described an
algorithm (ZTRACE) for recovering the real-space density field, the
peculiar velocities and the initial conditions from a redshift
catalogue with known selection function.  ZTRACE is based on a
self-consistent solution of the Zel'dovich approximation: it finds,
with an iterative scheme, the set of initial conditions which evolve
into the observed redshift-space density field, under the assumption
that galaxies trace mass.  The algorithm has been tested using N-body
simulations.  When the density field is smoothed with a
mass-preserving adaptive scheme, the initial conditions and their PDF
are reconstructed in an unbiased way, with the exception of the high
peaks whose density contrast, linearly extrapolated to the present
time, is $\ga 1$.  These have already gone into the highly non-linear
regime.  As shown by ME99, the quality of the reconstruction is almost
independent of the assumed background cosmology.

In this paper we apply the ZTRACE algorithm to the PSCz catalogue.  In
this way we recover the initial conditions of our local Universe on
scales as small as $\sim$5 \mpc, thus pushing the reconstruction to
the limit beyond which high non-linearity prevents any recovery.  Here
we focus on the reconstructed PDF of the initial conditions, testing
its Gaussian shape; future work will address the use of the
reconstructed initial conditions to simulate our local Universe.
Section 2 describes the PSCz catalogue and the construction of the
redshift-space density field.  Section 3 gives a brief description of
ZTRACE, and describes the results of its application to the PSCz
catalogue.  In Section 4 we analyze some sources of bias in the
reconstructed initial PDF, namely the high-density tail, the effect of
galaxy bias and the errors in the selection function.  Section 5
describes the reconstructed initial PDF of the PSCz catalogue, and
Section 6 gives the conclusions.

Throughout this paper distances are given in \mpc, with $h=H_0$/(100
km/s/Mpc).

\section{The PSCz catalogue}

For this analysis we use the recently completed PSCz catalogue, which
is described in detail in Saunders et al. (2000).  The catalogue is
based on the IRAS PSC (Joint IRAS Science Working Group 1988), and
comprises all galaxies with 60$\mu$ fluxes $>0.6$ Jy.  As the PSC is
not complete to 0.6 Jy in 2HCON areas (Beichman et al. 1988), the
1HCON detections recovered from the PSC Reject file were added to the
catalogue.  Stars, other galactic sources and AGNs were eliminated by
means of colour criteria, while residual contamination was eliminated
by a combination of optical identifications, examination of the raw
IRAS addscan profiles, examination of the Simbad database and, where
still unclear, spectroscopic analysis.  Redshifts for $\sim$ 4600
galaxies not belonging to the QDOT (Lawrence et al. 1999) or the 1.2
Jy (Fisher et al. 1995) surveys, and not available in literature, were
observed with several telescopes.  Faint galaxies, with $B_J>19.5$,
were not observed; most of these objects are galaxies with $z>0.1$,
and their omission is irrelevant for the present analysis.

The final catalogue is 98 percent complete in redshift over 84 percent of the sky.
The missing zones are the Galactic Plane (where the extinction is
$A_V>1.42$ mag), the areas covered by the LMC and SMC, and two small
strips not observed by IRAS.  Fig. 1 shows the Aitoff projected
catalogue, with the uncovered zones masked in grey.  The PSCz
catalogue has already been used in other works on the topology of the
large-scale structure (Canavezes et al. 1998), on the cosmological
dipole (Rowan-Robinson et al. 1999; Schmoldt et al. 1999a), on the
beta parameter (Schmoldt et al. 1999b), on the power spectrum of
fluctuations (Sutherland et al. 1999), on the redshift distortions
(Tadros et al. 1999) and on the peculiar velocity field (Branchini et
al.  1999). Comparisons with other galaxy and galaxy cluster
catalogues were presented by Seaborne et al. (1999) and Plionis et
al. (1999).  Reconstruction analyses were presented by Sharpe et
al. (2000), based on the least-action principle, and by the already
mentioned paper by Narayanan et al. (1999).

\begin{figure}
\centerline{\psfig{figure=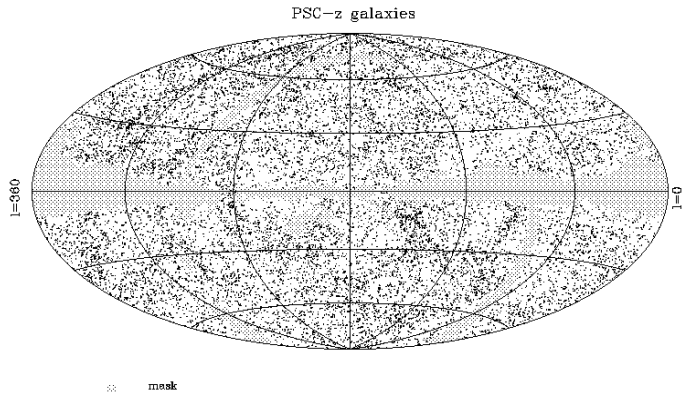,angle=-90,width=10cm}}
\caption{The PSCz catalogue in galactic coordinates. The regions
uncovered by the IRAS PSC are masked in grey.}
\end{figure}

To calculate the galaxy density field from the PSCz catalogue it is
necessary to know the selection function, which gives the number
density of objects bright enough to be included in the catalogue, as a
function of distance.  The selection function is well fitted by the
following functional form:

\be
\Phi(r)=\Phi_*\frac{(r/r_*)^{1-\alpha}}{(1+(r/r_*)^\gamma)^{\beta/\gamma}},
\label{eq:selfun} \ee

\noindent
where $r = cz/H_0$ is the redshift distance, but the volume is
corrected for relativistic effects (assuming an Einstein-de Sitter
Universe).  As ZTRACE applies in a frame comoving with the background,
we use redshifts in the CMB frame.  A maximum likehood fit, performed
with the method of Mann, Saunders \& Taylor (1996) and used in
Saunders et al. (2000), gives the following best values for the
parameters: $\Phi_*=0.0224\ h^3Mpc^{-3}$, $r_*=64.10$ \mpc,
$\alpha=1.41$, $\beta=4.32$ and $\gamma=1.39$.  The resulting
selection function is only marginally different from that calculated
in the Local Group frame.  The effects of uncertainties in the
selection function will be addressed in Section 4.3.

\begin{figure*}
\centerline{\psfig{figure=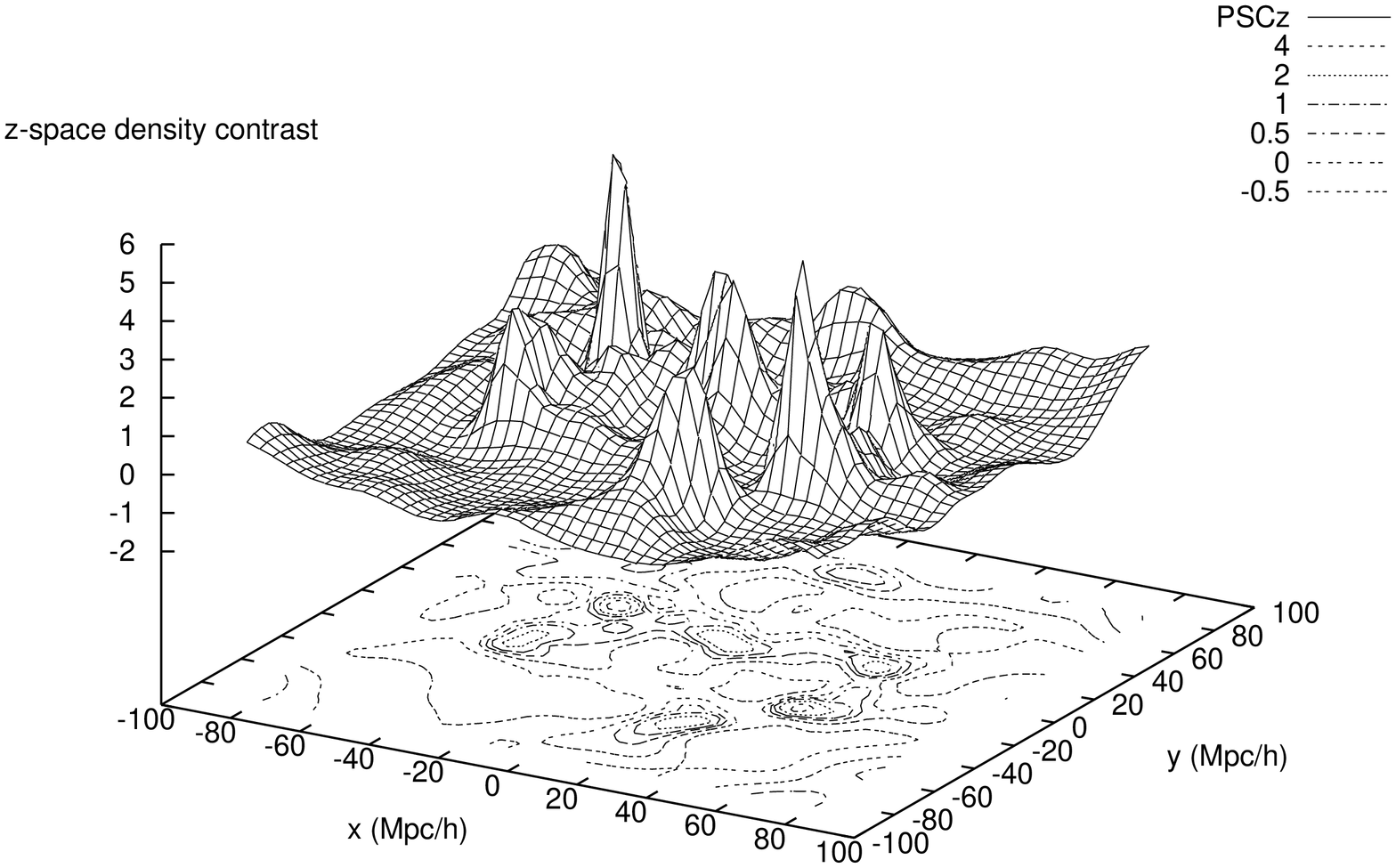,angle=-90,width=16cm}}
\centerline{\psfig{figure=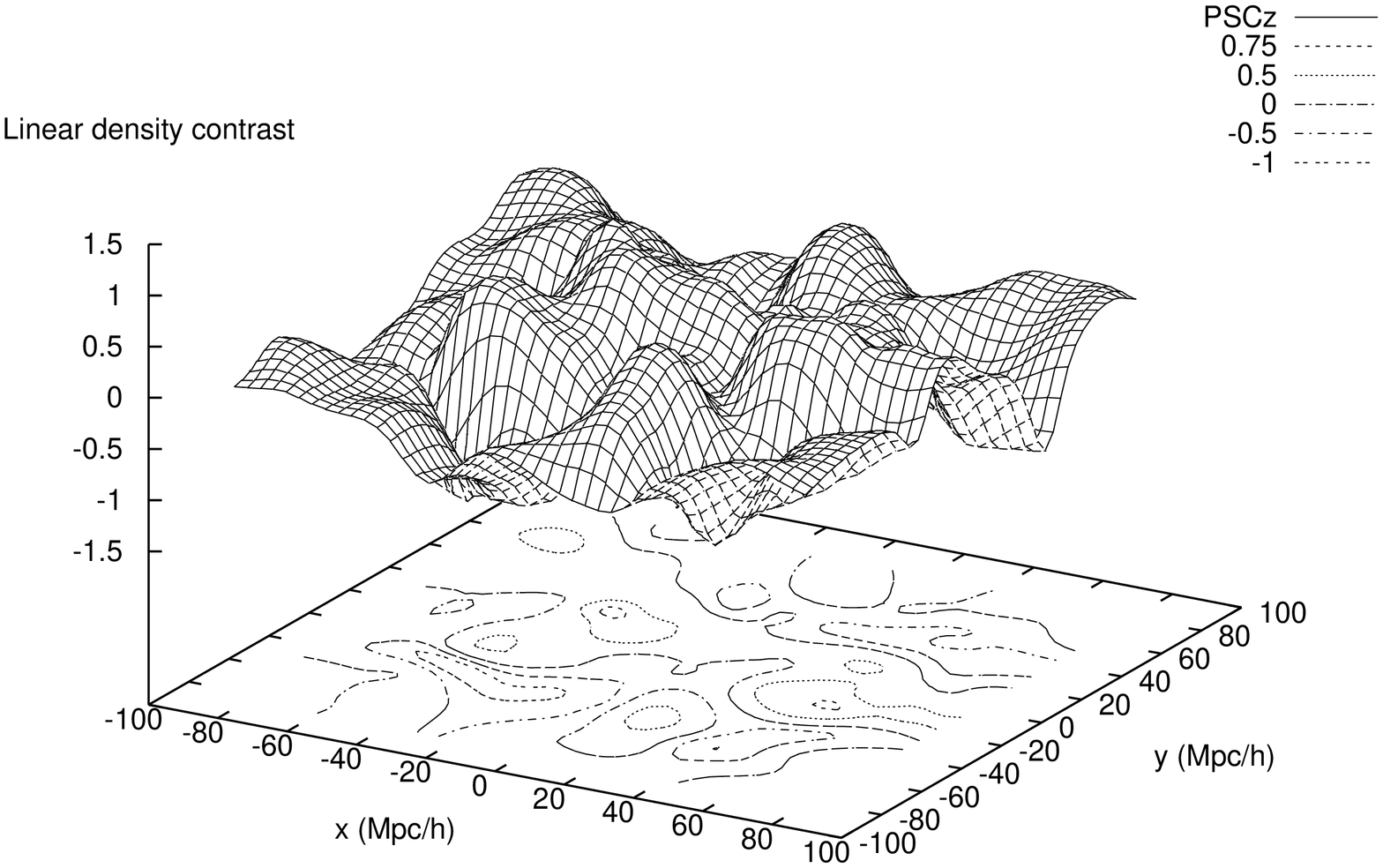,angle=-90,width=16cm}}
\caption{(a) Density field from the IRAS PSCz catalogue, adaptively
smoothed over 7.6 \mpc\ within 80 \mpc. The Super-Galactic Plane is
shown.  The peaks correspond to well known superclusters; see text for
details. (b) Initial density field, linearly extrapolated to the
present time.}
\end{figure*}

\section{Application of ZTRACE to the PSCz catalogue}

ZTRACE is based on a self-consistent solution of the Zel'dovich (1970)
approximation.  In the Lagrangian picture of fluid dynamics a mass
element with initial (Lagrangian) comoving position \q\ is displaced
to the final (Eulerian) position \x\ through a displacement field \S:

\be \mx(\mq,t)=\mq+\mS(\mq,t). \label{eq:map} \ee

\noindent 
The map \S\ is the dynamical variable in this approach.  For small
displacements, and for irrotational fluids, the first-order growing
mode gives the Zel'dovich (1970) approximation:

\be \mx(\mq,t)=\mq-D(t)\nabla\varphi(\mq). \label{eq:zel} \ee

\noindent
Here $D(t)$ is the linear growing mode (see, e.g., Peebles 1980); for
an Einstein-de Sitter universe it is equal to the scale factor $a(t)$.
The growing mode is normalized so that $D(t_0)=1$, where $t_0$ is the
present time.  $\varphi(\mq)$ is a rescaled version of the peculiar
gravitational potential, such that, at an initial time $t_i$,
$\nabla^2 \varphi= \delta(\mq,t_i)/ D(t_i)$.  The density contrast
$\delta$ is defined as usual, $\delta(\mx)= (\varrho(\mx)-
\bar\varrho)/ \bar\varrho$, where $\bar\varrho$ is the background
density.  It is useful to define the quantity:

\be \delta_l(\mq) \equiv \delta(\mq,t_i)/ D(t_i). \label{eq:lin} \ee 

\noindent
$\delta_l$ is constant in linear theory, and is equal to the initial
density contrast linearly rescaled to the present time.  This quantity
will be called the linear density contrast in the following.

In a redshift survey, the radial coordinate is the redshift distance,
not the real distance.  It is possible to generalize the Zel'dovich
map, Eq.~\ref{eq:zel}, to the redshift space \s:

\be \ms(\mq,t) = \mq - D(t)\left[\nabla\varphi(\mq)+f(\Omega)
\left(\nabla\varphi(\mq) \cdot \hat{\mx}\right) \hat{\mx}\right].
\label{eq:redshift} \ee

\noindent
Here $\hat{\mx}$ is the versor of \x, and $f(\Omega)=d\ln D/d \ln a$
is usually approximated by $f(\Omega)\simeq \Omega^{0.6}$ (Peebles
1980), or by $f(\Omega,\Omega_\Lambda) \simeq \Omega^{0.6}
+\Omega_\Lambda (1+\Omega/2)/70$ (Lahav et al. 1991) if the
cosmological constant is non-zero.  Notice that the last term in the
right-hand-side of Eq.~\ref{eq:redshift} is the peculiar velocity
along the line of sight.

The density contrast follows from the map in Eq.~\ref{eq:zel} or
Eq.~\ref{eq:redshift} through:

\be 1+\delta(\mq)=[\det(\delta^K_{ab}+S_{a,b})]^{-1}. \label{eq:dens} \ee

\noindent
Here $\delta^K_{ab}$ is the Kronecker tensor, and $S_{a,b}$ is the
derivative of the map \S.  ZTRACE is designed to invert the non-linear
and non-local set of equations which connect the initial conditions to
the redshift-space density, under the hypothesis of laminar flow,
i.e. when the Zel'dovich map in redshift space, Eq.~\ref{eq:redshift},
is still single-valued.  Multi-valued regions appear as soon as
different mass elements which lie along the same line of sight are
observed at the same redshift.  This happens when the perturbation
decouples from the Hubble flow, giving rise to multi-stream regions in
the distance-redshift relation (often referred to as ``triple-valued
regions'').  The linear density contrast has a value of order one in
such regions; ZTRACE is not able to reconstruct the high peaks of the
density distribution.

ME99 describe in detail those techniques used to help the convergence
of ZTRACE for smoothed density fields with standard deviations
$\sigma$ up to 0.7\footnote{The tests performed by ME99 are limited to
$\sigma\le 0.7$.  Anyway, for larger $\sigma$-values the distortion on
the positive tail of the initial PDF becomes very strong.}.  The
application of ZTRACE to the PSCz redshift catalogue can be summarized
as follows.

(i) The masked regions are filled with a synthetic catalogue, so as to
produce the galaxy density field over a large and connected spherical
volume.  We have used the procedure developed by Yahil et al. (1991)
as modified by Branchini et al. (1999).  According to this procedure,
the zone of avoidance of the Galactic disk, with galactic latitude
$|b|<8$ degrees, is divided in longitude and redshift bins.  The
galaxy number density in each bin is obtained by interpolating between
those of the two 8 degree stripes above and below the bin.  A
synthetic galaxy catalogue is then obtained by randomly sampling the
density field within the bins.  In this way, the radial distribution
follows the same PSCz selection function.  The other masked regions of
the sky which are outside the zone of avoidance are filled at random,
according to the PSCz selection function.  This algorithm is
considered a better guess with respect to filling at random the masked
regions, as it tends not to break the coherent structures, such as the
Perseus-Pisces supercluster, which happen to cross the Galactic
Plane. As a further test, we have also applied the lognormal
interpolation procedure described in Saunders et al. (1999); the
results are very similar and are not described here.  The
reconstruction method is robust with respect to the interpolation
algorithm because only a small fraction of the sky is masked.

(ii) The relaxed groups are collapsed into spheroids.  The effects of
the highly non-linear dynamics are apparent in the redshift space as
elongated structures along the line of sight, often referred to as
`fingers of God'.  As pointed out by Gramann, Cen \& Gott (1994), a
reconstruction scheme improves if such non-linear structures are
collapsed before application of the algorithm.  In ME99 the collapsing
procedure was found to give a modest increase in the performance of
ZTRACE.  Following ME99, we have found the groups in the PSCz
catalogue by applying a standard friends-of-friends algorithm, with
radial and tangential linking lengths of 3 and 0.5 \mpc.

\begin{figure*}
\centerline{ \psfig{figure=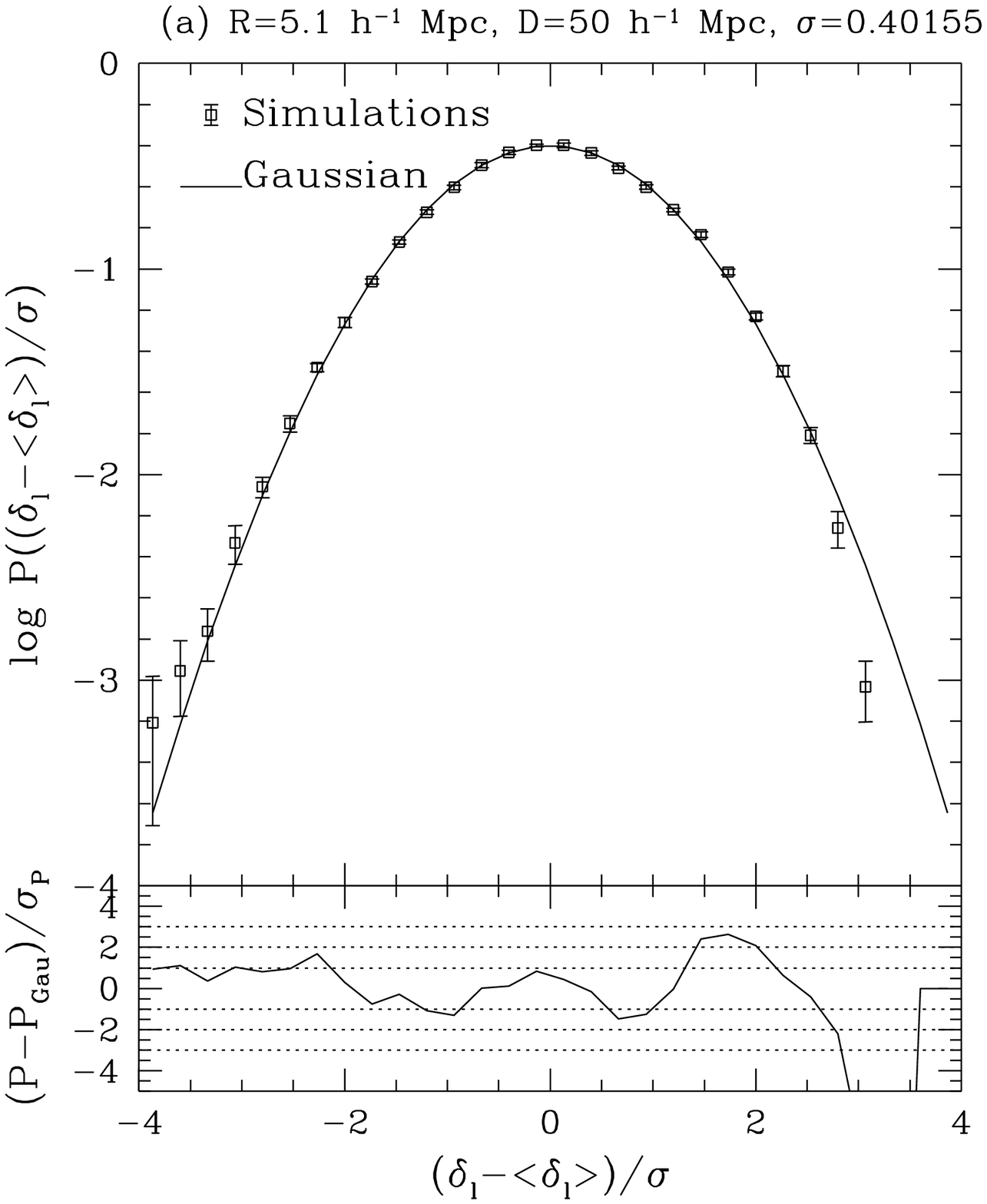,width=10cm}
\psfig{figure=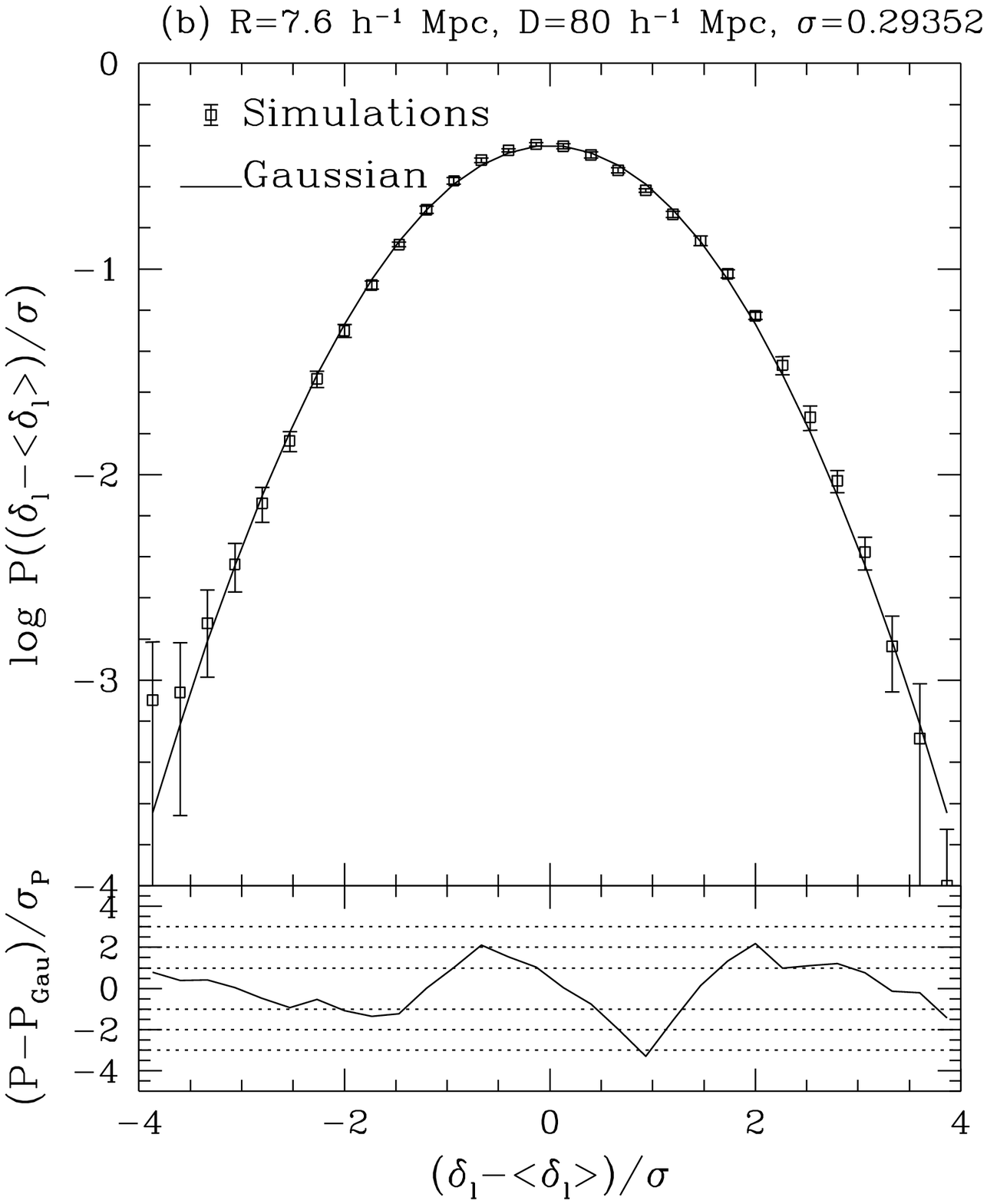,width=10cm}}
\caption{Reconstructed initial PDFs, averaged over 10 simulated
catalogues.  The errorbars shown are the square root of the variance
of the mean among the realizations.  The true Gaussian shape is also
shown.  (a) Smoothing $R=5.1$ \mpc\ within $D=50$ \mpc. (b) Smoothing
$R=7.6$ \mpc\ within $D=80$ \mpc.  The lower panels show the
significance of the difference between the simulated and true initial
PDFs.  The lines at $\pm 1\sigma$, $\pm 2\sigma$ and $\pm 3\sigma$ are
shown for clarity.}
\end{figure*}

(iii) The catalogue is smoothed to generate a density field in
redshift space.  The performance of ZTRACE improves considerably if
the density field is obtained with an adaptive, mass-preserving
filtering of the galaxy catalogue.  This is similar to a smoothing in
the Lagrangian space, and correctly takes into account that
overdensities come from larger patches which have contracted, while
the opposite is true for underdensities.  Following ME99, we have
constructed the redshift-space galaxy density field by smoothing the
PSCz point distribution (with the addition of the synthetic catalogue
to fill the masked regions) with the adaptive smoothing code of
Springel et al. (1998).  This code requires the definition of a
reference smoothing radius, which corresponds to the case of null
density contrast; the adaptive refinements are then performed so that
the smoothing filter always contains the same number of objects.  As
in ME99, the reference smoothing radius $R$ is held constant within a
given distance $D$, to a value such that 10 galaxies are on average
contained in the filter at that distance.  At larger distances, the
reference smoothing radius is increased according to the selection
function, so that 10 galaxies are contained on average within the
filter.  In this way, the outer parts that are more and more severely
smoothed are used by ZTRACE to give the external tides to the inner
region.

The density field is calculated on a $64^3$ grid of size 240 \mpc.
Fig. 2a shows the obtained density field in Super-Galactic
coordinates, in the case $R$=7.6 \mpc\ within $D$=80 \mpc; the severe
smoothing of the outer regions is apparent.  The high peaks in the
density distribution correspond to well-known structures as the Local
Supercluster [at the position $(x,y)\sim (0,0)$], the Great Attractor
region [the two peaks at $(x,y)\sim (-40,-20)$ and $(-40,20)$], the
Perseus-Pisces supercluster [the two peaks at $(x,y)\sim(20,-50)$ and
$(40,-20)$] and its appendix beyond the Galactic Plane [the
Camelopardalis supercluster, at $(x,y)\sim (60,10)$], and the Coma
supercluster [at $(x,y)\sim (0,80)$] (see, e.g., Branchini et al.
1999).

(iv) The input density field is padded in the volume outside the
largest inscribed sphere (of radius 120 \mpc), so as to have periodic
boundary conditions.  This is done so that we can apply FFT's to solve
the equations.  ZTRACE is applied to the field; convergence is
achieved within $\sim$12 iterations.  Fig. 2b shows the resulting
initial conditions in the comoving real space (in terms of the linear
density contrast $\delta_l$, Eq.~\ref{eq:lin}).  (Notice that the
linear density contrast is not bound to be $\ge -1$).  The main
superclusters are originated from high peaks in the initial
conditions.  These peaks appear flattened because of the inability,
discussed above, of ZTRACE to reconstruct the high-density regions.
The different statistical properties of the initial conditions, with
respect to the density field in real and redshift space, are apparent.

(v) The PDF of the initial conditions is calculated.  In this analysis
only the grid points within the distance $D$ are considered.  The
points in the zone of avoidance ($|b|<8^\circ$) are excluded, so as
the central 10 \mpc\ sphere.  This last exclusion is motivated by the
inability of ZTRACE to converge in the neighborhood of the observer;
the numerical solution tends to oscillate at the origin, where the map
given in Eq.~\ref{eq:redshift} is singular.  To avoid this the density
field is damped over $\sim$5 \mpc.  The exclusion of the central
region is not a problem as its volume is small, and the PSCz is known
to be incomplete at such small distances (Rowan-Robinson et al. 1991).

\section{errors and Biases in the reconstruction of the initial PDF.}

The recovery of the initial PDF can be influenced by high
non-linearities, by galaxy bias and by errors in the selection
function.  As shown in ME99, a wrong assumption on the background
cosmology does not influence significantly the shape of the initial
PDF.

\subsection{The high-density tail}

As mentioned above, ZTRACE is not able to recover the high peaks,
which have decoupled from the Hubble flow.  This induces a bias in the
initial PDF, whose high density tail is suppressed.  The
underestimated high-density points influence the initial PDF also at
moderate densities.  It is useful to quantify this effect before
addressing the Gaussian nature of the initial PDF reconstructed from
the PSCz catalogue.  To this aim, we have extracted mock PSCz
catalogues from an N-body simulation of a standard CDM Universe (with
$\Omega=1$ and $\Gamma\equiv\Omega h= 0.5$) with normalization
$\sigma_8=0.7$.  The simulation, performed with the Hydra code
(Couchman, Thomas \& Pearce 1995), has already been presented in ME99.
It is performed with $128^3$ dark matter particles in a box of length
240 \mpc\ with a $256^3$ base mesh with adaptive refinements.  As the
volume used in the reconstruction is much smaller than the total
volume of the simulation, it is convenient to extract more than one
simulated PSCz catalogue from the simulation.  We have extracted 10
catalogues, centred on random points, under the assumptions that
galaxies trace mass and applying the PSCz selection function (in real
space).  It has been checked in ME99 that the selection in real space
with a selection function determined in redshift space does not
influence the result.  The smoothing radius has been set to $R=5.1$
\mpc\ (constant within $D=50$ \mpc) or $R=7.6$ \mpc\ (constant within
$D=80$ \mpc).  When the smoothing is held constant within $D=50$ \mpc,
the volume actually used in the reconstruction is 3.8 percent of the
total volume of the simulation, so that the different realizations are
accurately statistically independent.  When the smoothing is held
constant within $D=80$ \mpc, the volume used is 15.5 percent, and then
the 10 realizations are oversampling the volume.  This is
approximately taken into account by calculating the variance among the
various realizations as if they were $1/0.155\simeq 6.45$ instead of
10.

For each simulated catalogue, the galaxies in the PSCz masked area are
removed, then replaced using the same algorithm of Branchini et
al. (1999) discussed above.  We have checked that the masking
procedure introduces noise but no bias in the reconstruction.  The
reconstruction proceeds as described in Section 3.

Because of sample variance, the reconstructed initial PDFs differ in
general in mean ($\langle\delta_l \rangle$) and standard deviation
($\sigma_l$).  To construct an average initial PDF it is then useful
to consider the PDF of the quantity $(\delta_l-\langle\delta_l
\rangle)/\sigma_l$.  All the distributions are binned in the same way,
and averages and variances are calculated for each bin.  Fig. 3 shows
the reconstructed initial PDFs, compared with the true Gaussian curve,
for the two smoothing radii used.  The upper panels show the
reconstructed initial PDFs, the errors are the square root of the
variance of the mean for the ten catalogues (calculated as if they
were 6.45 in the second case, see above).  The lower panels show the
significance of the difference between the reconstructed and true
initial PDF.  It is apparent that the true Gaussian shape is recovered
accurately, except in the high-density tail.

\begin{figure}
\centerline{\psfig{figure=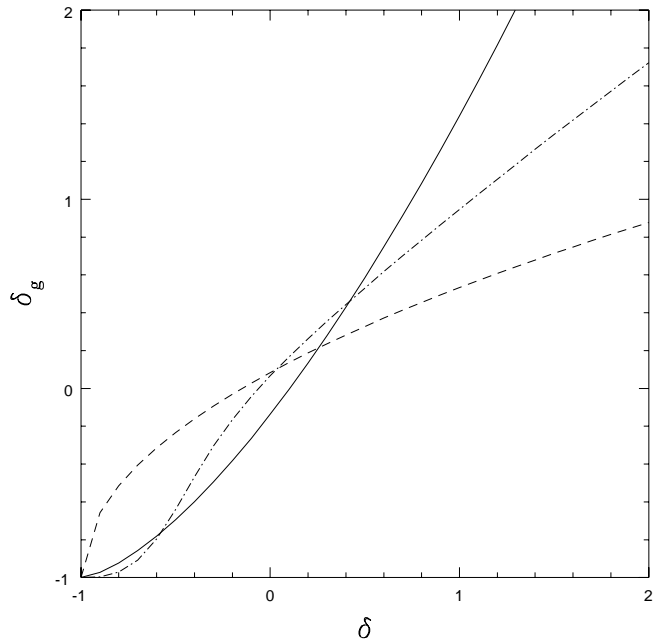,width=9cm}}
\caption{Bias functions used in the paper. Dashed line: anti-biased
power-law with $b_1=0.5$.  Continuous line: biased power-law with
$b_1=1.5$.  Dot-Dashed line: semi-analytical bias.}
\end{figure}

\subsection{The effect of galaxy bias.}

The simulated galaxy catalogues are created and reconstructed assuming
that galaxies trace mass.  This may not be true in practice, as
galaxies may be biased tracers of the underlying mass distribution.
As mentioned in the Introduction, galaxy bias may induce non-Gaussian
distortions in the initial PDF, when this is reconstructed simply
assuming that galaxies trace mass.  To quantify this effect, it is
convenient to apply ZTRACE, under the assumption that galaxies trace
mass, to biased galaxy catalogues that are simulated assuming some
specific bias model.

We have used a (deterministic, i.e. non-stochastic) power-law local
bias scheme, already suggested, e.g., by Nusser et al. (1995) and by
Lahav \& Dekel (1999):

\be 1 + \delta_g(\mx) = (1+b_0) (1 + \delta(\mx))^{b_1}. \label{eq:bias} \ee

\noindent
Here $\delta_g$ is the galaxy density contrast.  This scheme respects
the condition $\delta_g>-1$.  For small density values this scheme
reduces to the linear bias, $\delta_g =(1+b_0) b_1 \delta$.  $b_0$ is
constrained so that $\langle \delta_g \rangle = 0$.  In the following
we will consider the anti-biased case $b_1=0.5$ and $b_0=0.084$, and
the biased case $b_1=1.5$, and $b0=-0.137$.  Fig. 4 shows the two
bias curves.

\begin{figure*}
\centerline{
\psfig{figure=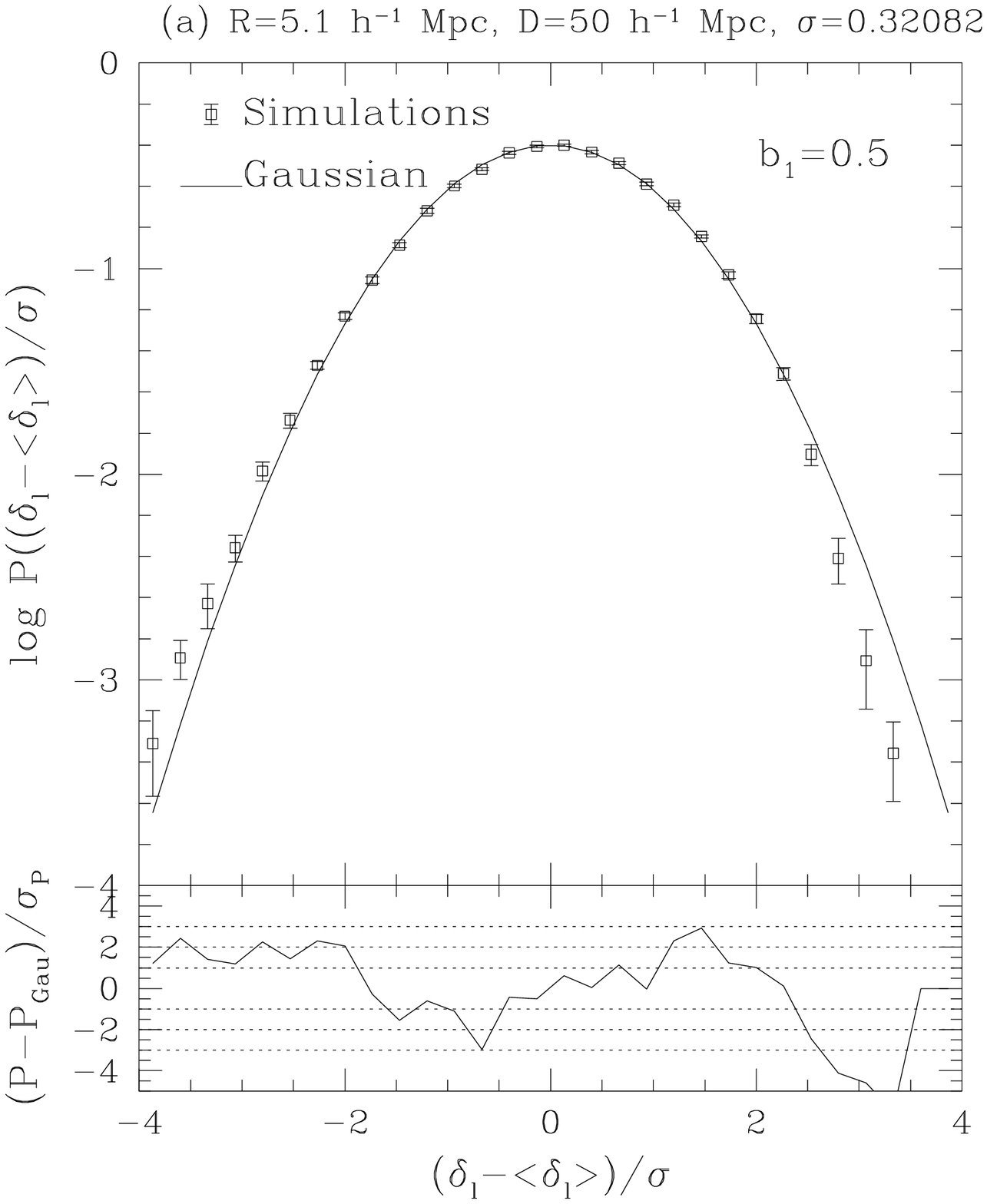,width=7cm}
\psfig{figure=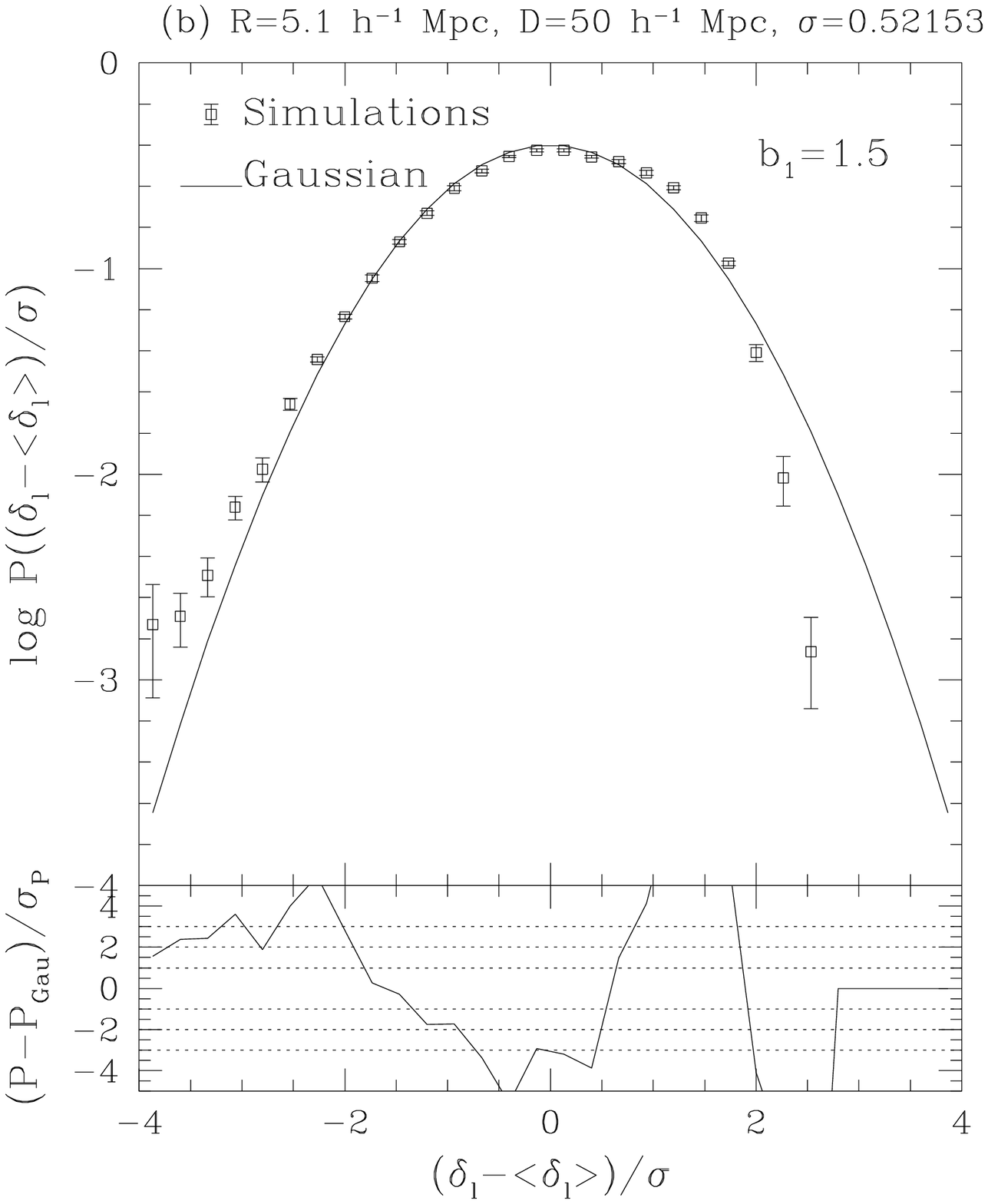,width=7cm}
\psfig{figure=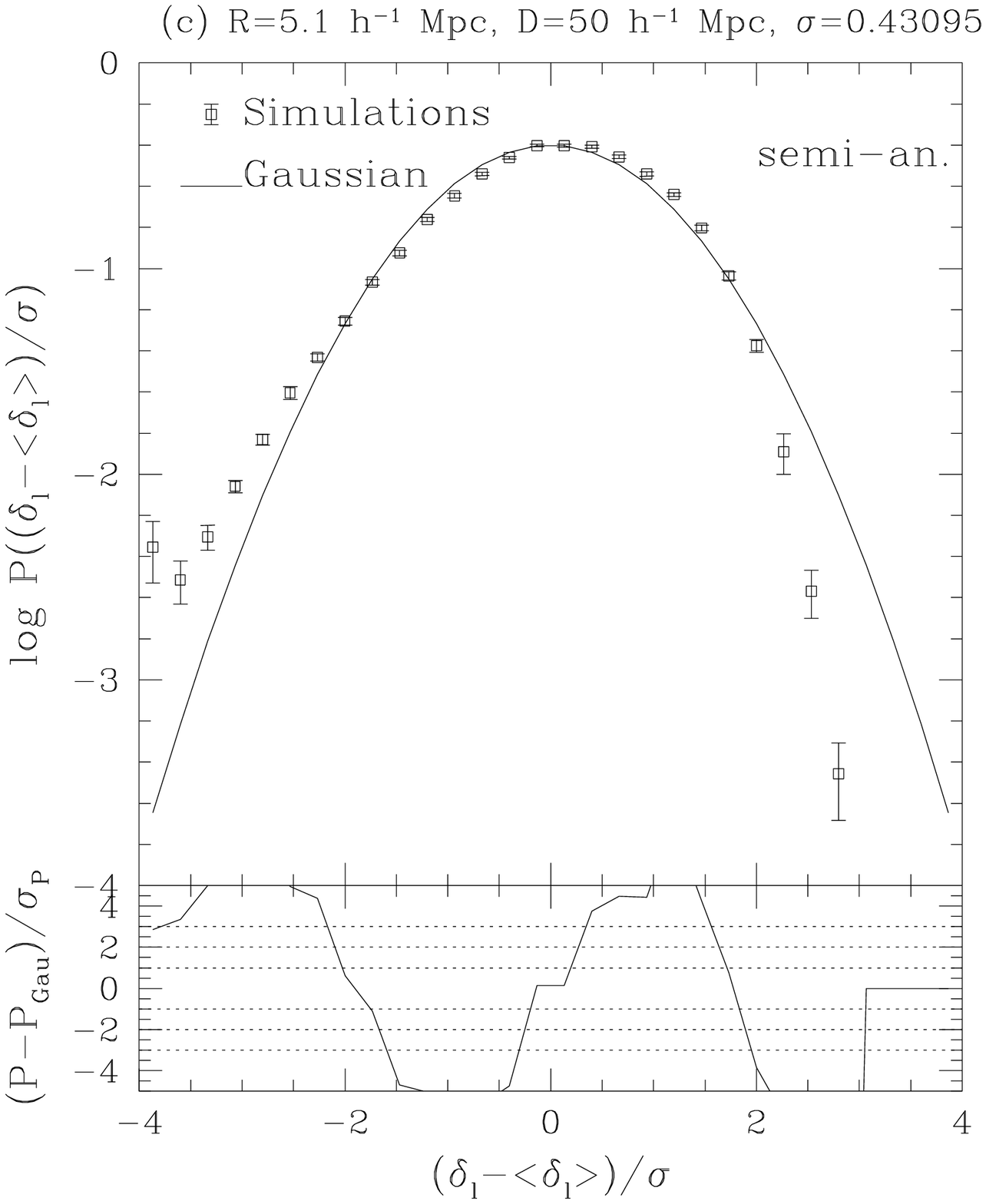,width=7cm}}
\centerline{
\psfig{figure=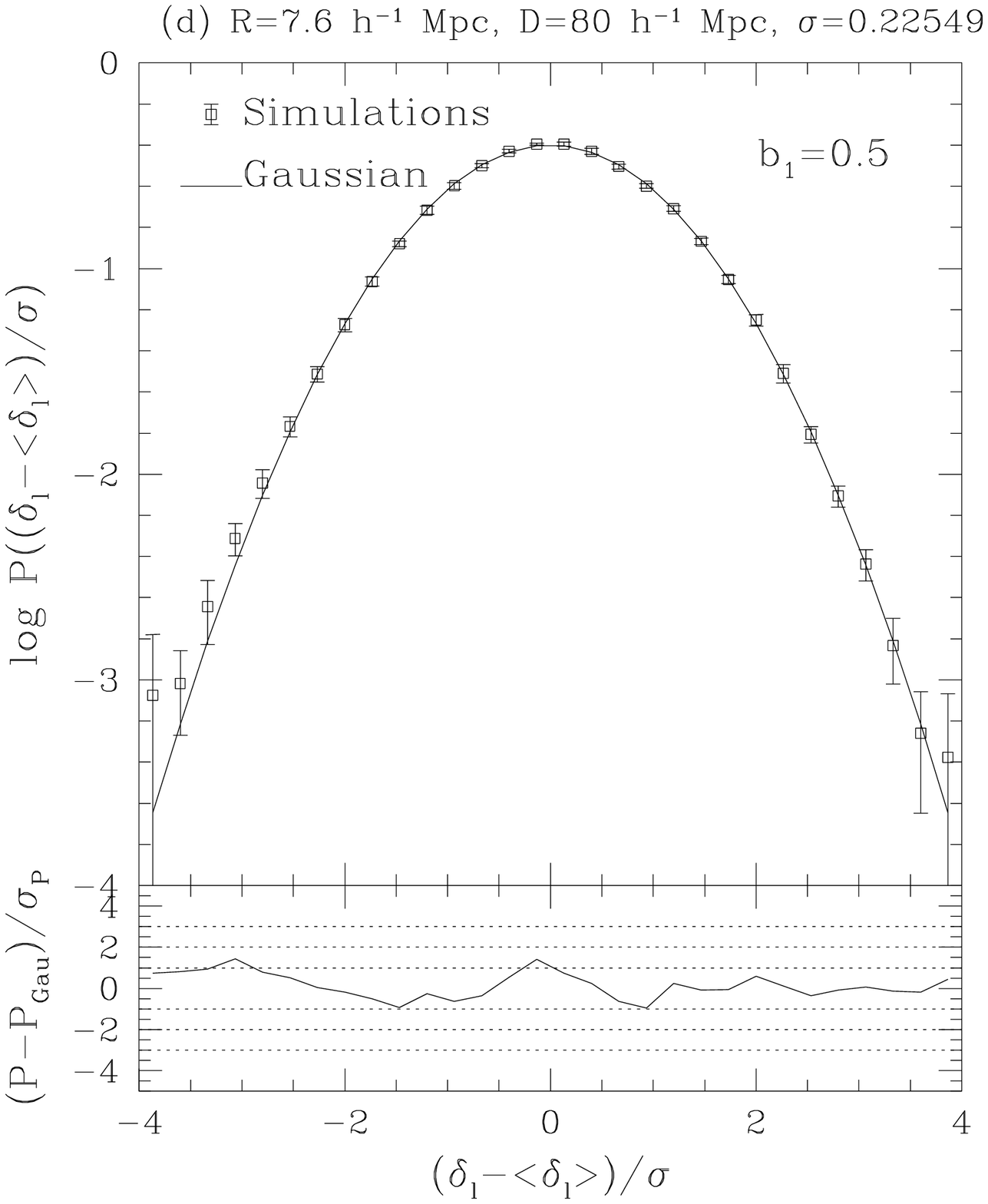,width=7cm}
\psfig{figure=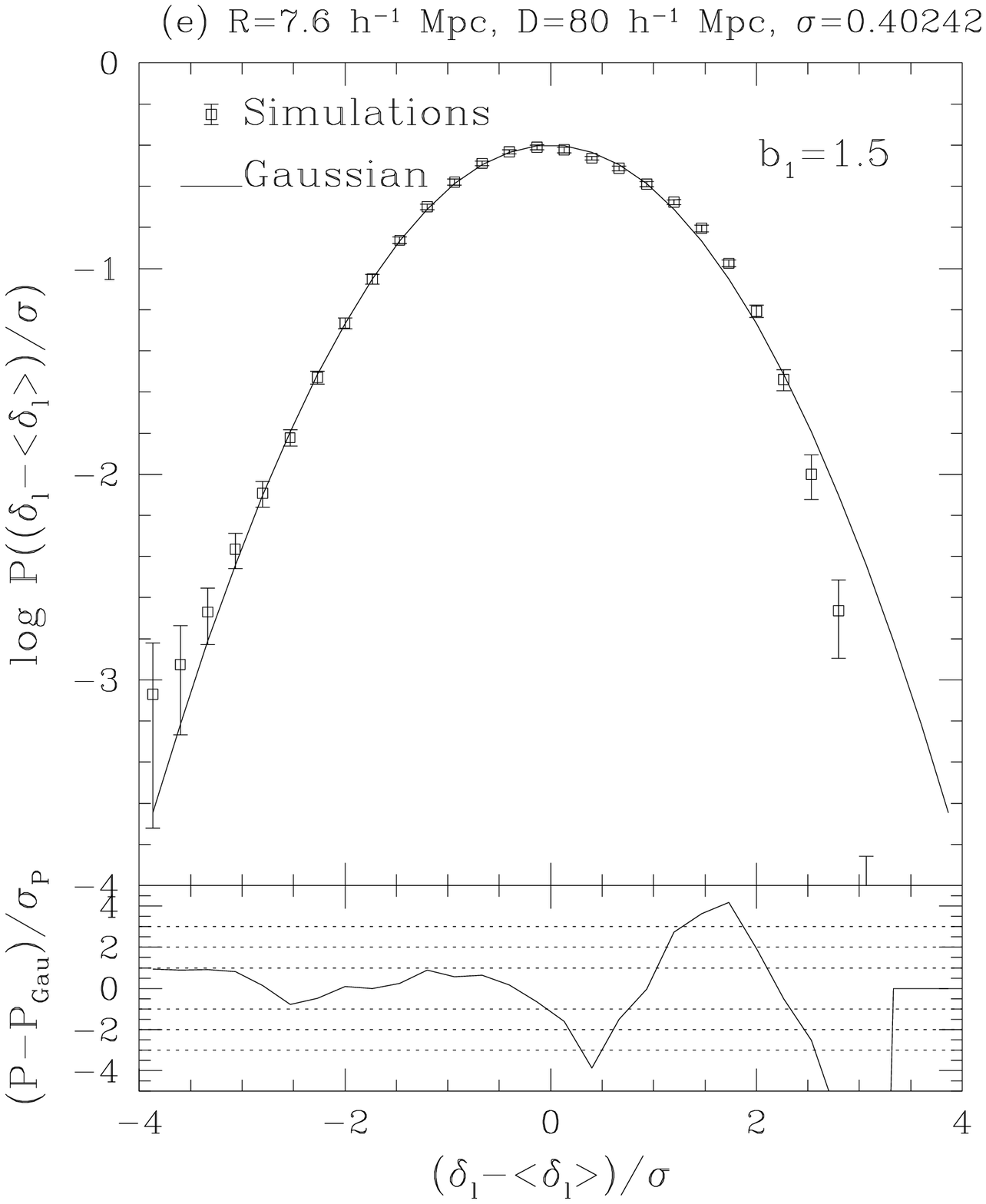,width=7cm}
\psfig{figure=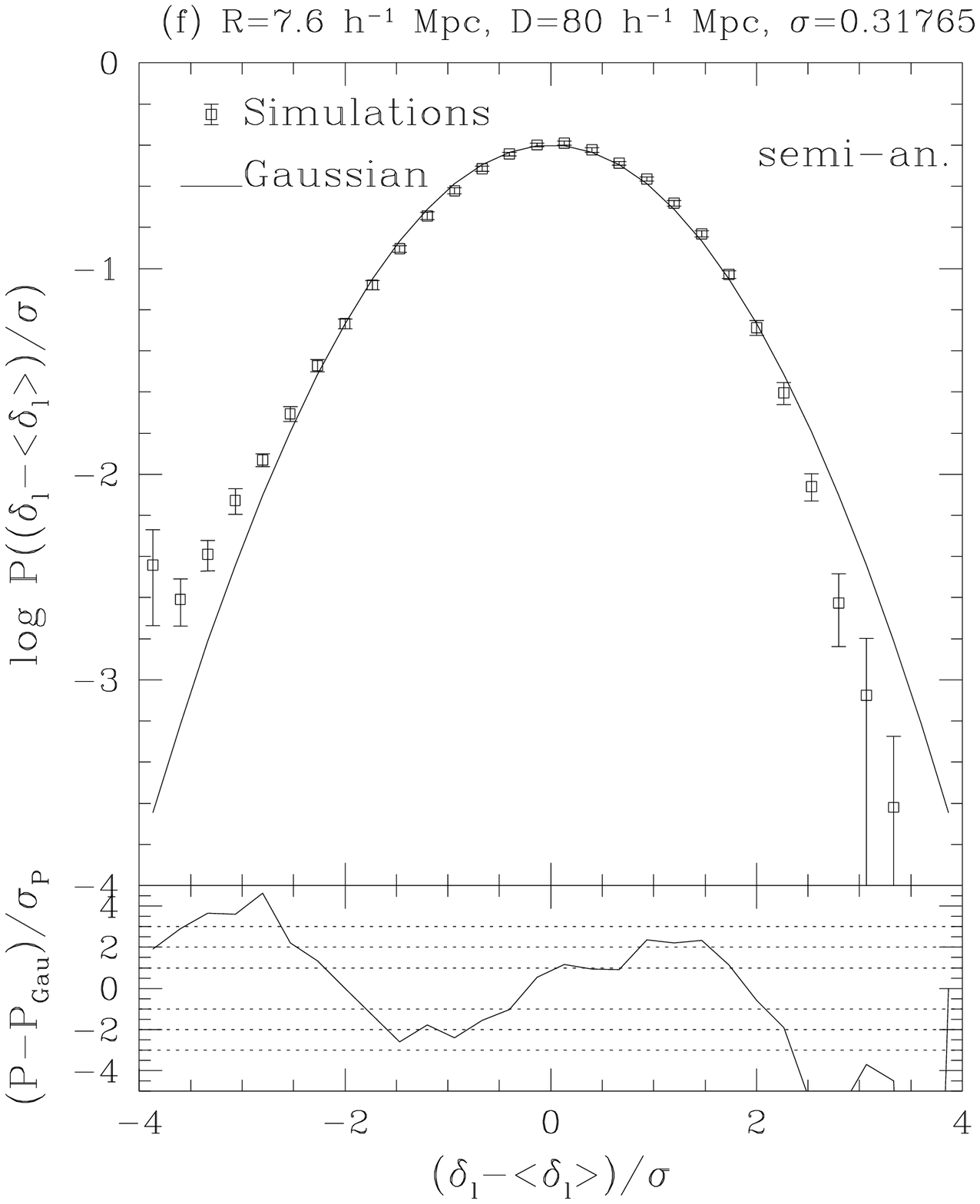,width=7cm}}
\caption{As in Fig. 3: reconstructed initial PDFs, averaged over 10
simulated catalogues, for biased catalogues, with $b_1=0.5$ (a,d),
$b_1=1.5$ (b,e) or semi-analytic bias (c,f).  The smoothing radius is
set to $R=5.1$ \mpc\ within $D=50$ \mpc\ (a,b,c) or $R=7.6$ \mpc\
within $D=80$ \mpc\ (d,e,f).}
\end{figure*}

A more physically motivated bias scheme can be obtained by using
semi-analytic galaxy formation models (see, e.g., Benson et al. 1999;
Somerville et al. 1999).  In these models, the assembly of dark matter
halos is described by means of N-body simulations, while the physics
of the baryonic component is inserted in the halos through a set of
simple analytic rules.  Galaxy bias can then be obtained by comparing
the density of galaxies with the density of dark matter.  Narayanan et
al. (1999) give an approximate bias function for the IRAS-like
galaxies of Benson et al. (1999):

\be 1+\delta_g(\mx) = A (1 + \delta(\mx))^{\alpha} \left[ C + 
(1 + \delta(\mx))^{(\alpha-\beta)/\gamma}\right]^{-\gamma}.
\label{eq:benson} \ee

\noindent
Here $\alpha=2.9$ and $\beta=0.825$ are the asymptotic logarithmic
slopes at small and large densities, while $A=1.1$, $C=0.08$ and
$\gamma=0.4$.  This bias function is shown in Fig. 4.  Compared to the
simple power-law curves, it behaves as in the biased case in
underdensities, but it is qualitatively more similar to the unbiased
case in overdensities.  We use the semi-analytic bias scheme as an
exercise to quantify the distortion of the initial PDF expected from a
more sophisticated bias function.  It must be noticed, however, that
this bias function is relative to a flat Universe with
$\Omega_\Lambda=0.7$ and $\sigma_8=0.9$, and to a density field
smoothed with a top-hat filter with radius 3 \mpc, while we are
applying it to an Einstein-de Sitter universe and to a density field
Gaussian smoothed over 5 \mpc.

Following the same procedure as in Section 4.1, we have simulated
three sets of 10 catalogues, with the three bias functions discussed
above and shown in Fig. 4.  The bias schemes are applied to the
real-space density field of the simulation, smoothed with a Gaussian
filter of width 5 \mpc.  Fig. 5. shows the obtained average initial
PDFs; they are to be compared to those shown in Fig. 3.  The Gaussian
initial PDF is again well reconstructed in the ``anti-biased''
power-law case in which $b_1=0.5$, even though the low-density tail
shows a 2-$\sigma$ discrepancy when $R=5.1$ \mpc.  In the ``biased''
power-law case, that in which $b_1=1.5$ and especially the
semi-analytic case, significant distortions are apparent.  For $R=5.1$
\mpc, the tail at low densities is overestimated, a dip and a bump are
present respectively at $\delta_l/\sigma\sim 0$ and 1.5.  The same
features are still present in the case $R=7.6$, although they are less
significant.

\begin{figure}
\centerline{\psfig{figure=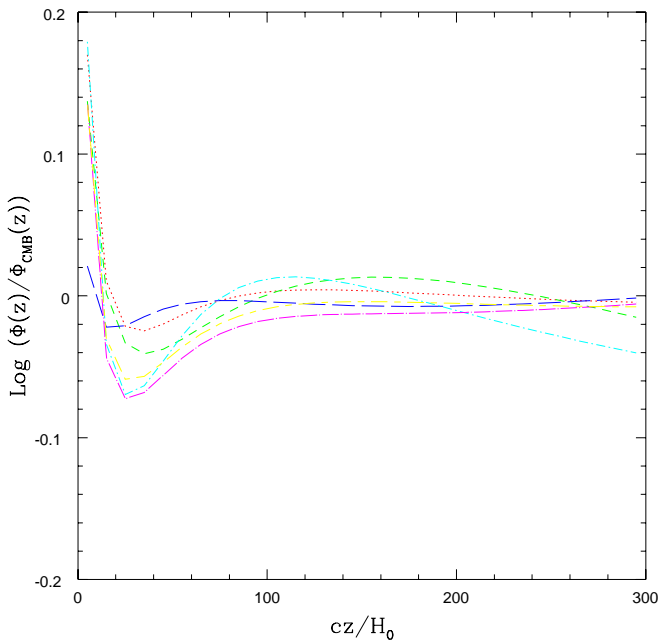,width=9cm}}
\caption{Residuals of the selection functions taken from ME99
(long-dashed), Sutherland et al. (1999) (dashed), Branchini et
al. (1999) (dot-dashed), Canavezes et al. (1999) (dot-long dashed),
Springel (1996) (short-long dashed), and determined by us in the Local
Group frame (dotted), with respect to the one calculated by us in the
CMB frame.}
\end{figure}

Fig. 5 shows that the application of ZTRACE with the no-bias
assumption can lead to small but significant distortions in the
reconstructed initial PDF.  However, these distortions are revealed by
averaging over ten realizations of the PSCz catalogue, but they are
not statistically significant for single biased realizations.  The
PSCz survey is not large enough to reveal such deviations, but larger
surveys such as the 2dF and SDSS surveys may be able to provide
constraints on non-linear bias schemes.  Of course this conclusion may
change if stronger bias functions are considered; however, evidence on
IRAS galaxies suggests that they are weakly biased with respect to the
matter density (see, e.g., Saunders et al. 1999).

\subsection{Errors in the selection function}

Another potential source of distortion on the initial PDF comes from
errors in the selection function.  Given the highly correlated nature
of the density field, a systematic error in the density estimate in a
spherical shell which happens to contain some large voids or clusters
can influence the shape of the initial PDF.  To test for this, we have
collected the selection functions (in the Local Group frame) used by
ME99, Sutherland et al. (1999), Canavezes et al. (1998), Branchini et
al.  (1999), and Springel (1996).  We have added to these our own
determination of the selection function in the Local Group frame,
performed as described in Section 2.  The differences between the
various functions, which are of order of 10 percent, are due to the
slightly different versions of the catalogue and mask used by the
various authors, and to differences in the fitting procedure.  Fig. 6
shows the residuals of these selection functions with respect to the
one used in this paper (Eq.~\ref{eq:selfun}).  Notice that all the
residuals show the same systematics because the selection functions
are relative to the Local Group frame.  Then, the effect of errors in
the selection function is overestimated in this analysis.

\begin{figure*}
\centerline{
\psfig{figure=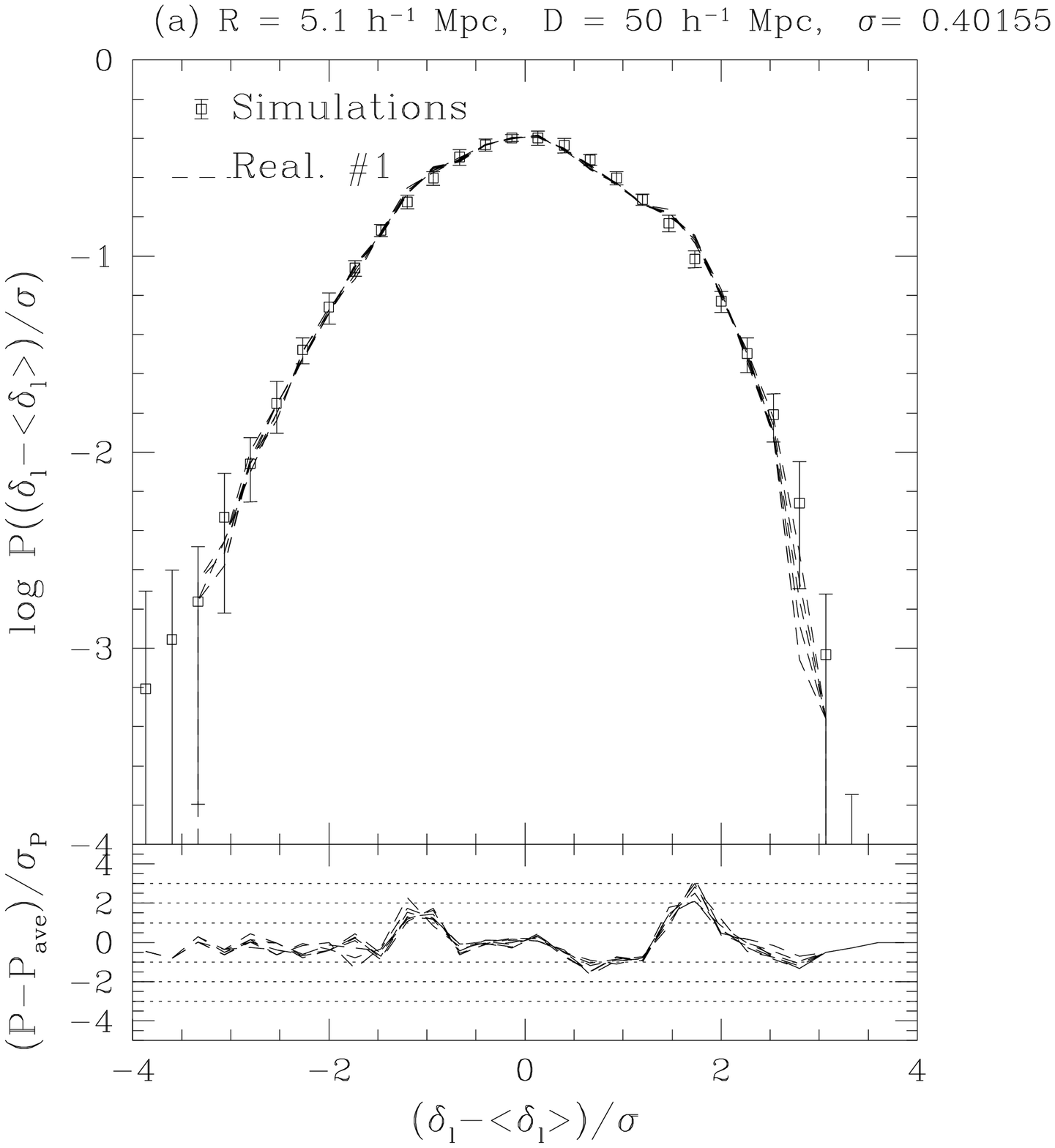,width=10cm}
\psfig{figure=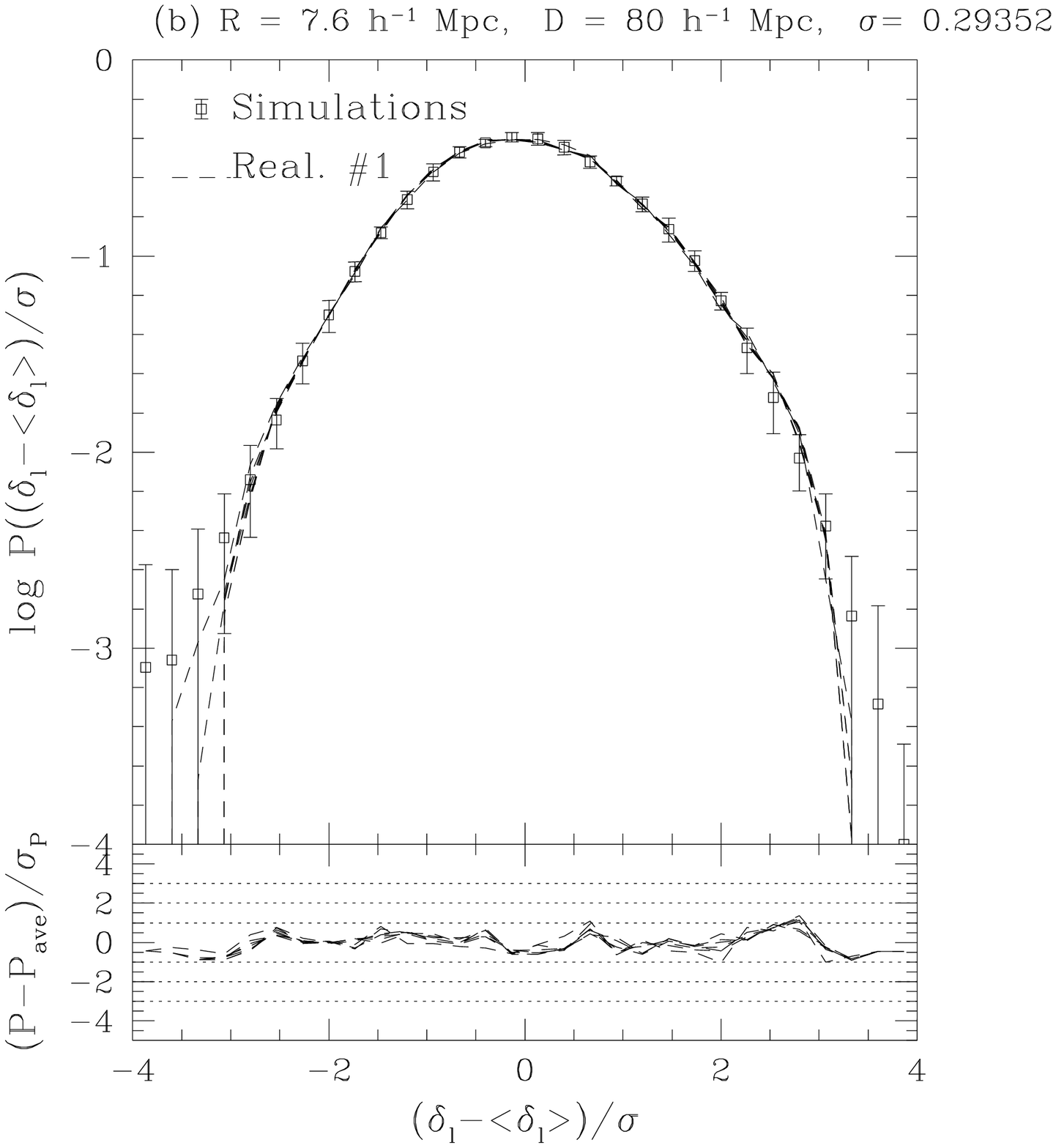,width=10cm}}
\caption{Reconstructed initial PDF for the first simulated (unbiased)
catalogue, compared to the average simulated initial PDF of Fig. 3.
The density field is reconstructed using either the correct selection
function or one of the six alternatives listed in the text.  The error
associated to the simulated initial PDFs is the square root of the
variance among the realizations.  The lower panels show the
significance of the disagreement between the given realization and the
average.}
\end{figure*}

The first simulated PSCz catalogue (with no bias) has been
reconstructed assuming either the correct or one of the six other
selection functions listed above.  Fig.  7 shows the seven
reconstructed initial PDFs, compared with the average reconstructed
initial PDF already shown in Fig. 3.  Differently from Fig. 3, the
errors shown are the (square root of the) variance between the various
realizations (instead of the variance of the mean).  This is the
correct error to use when assessing the difference between a single
realization and the average initial PDF.  Again, the lower panels show
the significance of the disagreement.  The various initial PDFs are
hardly distinguishable, no significant distortion is induced.  We
conclude that the uncertainty in the selection function does not
influence significantly the shape of the reconstructed initial PDF.

\section{The reconstructed initial PDF of the PSCz catalogue}

The PSCz catalogue has been adaptively smoothed according to the
smoothing strategy described in Section 3.  The reference radius has
been held constant, within $D=50$ and 80 \mpc, to the values of
$R=5.1$ and 7.6 \mpc.  Fig. 8 shows the reconstructed initial PDFs of
the PSCz catalogue compared with the results of the simulated
catalogues described in Section 3.1 (with no bias and CMB selection
function) and shown in Fig. 3.  The error associated to the simulated
initial PDFs is the square root of the variance between the
realizations.

The value of $\sigma_8$ for the PSCz density field turns out to be
about 0.8 (Sutherland et al. 1999; Seaborne et al. 1999), slightly
larger than the value 0.7 used in our test simulation.  As a
consequence, the variance of the reconstructed initial PDF is slightly
larger than the test one smoothed with the same radius.  This has some
effect on the spurious high-density cutoff of the PDF induced by
ZTRACE.  To compare reconstructed PDFs that have roughly the same
variance, the PSCz has been smoothed over $R=7.2$ and 10.1 \mpc\
within $D=75$ and 105 \mpc.  Fig. 9 shows the comparison with the same
PDFs of simulated catalogues as in Fig. 3 and 8.

The reconstructed initial PDF of the PSCz catalogue disagrees
significantly with the simulated one in Fig. 8a (smoothing radius
$R=5.1$ \mpc), where the standard deviation $\sigma$ of the two PDFs
that are compared are different.  Even in this case, significant
distortions are visible only on the positive-density side, for
$(\delta_l- \langle\delta_l\rangle) /\sigma\ga 1$.  In this range of
densities the influence of the high-density cut (which is different in
the two cases due to the difference in $\sigma$) is strong.  This
discrepancy vanishes in Fig. 9a, when comparing the reconstructed PDFs
of simulated and PSCz catalogues with similar variances.  Besides, the
distortions on the negative-density side, where we expect to find the
signature of galaxy bias (see Fig. 4), are never significant, except a
3-$\sigma$ discrepancy which is limited to one single bin.  Figs. 8b
and 9b show the same comparison for the larger smoothing radii.  In
this case, no strongly significant distortions are visible, although
some modest but persistent discrepancy at the 3-$\sigma$ level is
present both on the positive- and on the negative-density sides.

We conclude that there is no convincing evidence for non-Gaussianity
in the initial PDF reconstructed from the PSCz catalogue.  Notably,
the overall pattern of the residuals is similar to the distortions
induced by biased power-law or semi-analytic bias.  Anyway, the
statistical significance of these distortions is too marginal to draw
any conclusion.

The Gaussianity of the reconstructed inital PDF gives support to the
hypotheses of Gaussianity of initial conditions and of modest bias of
IRAS objects (as described in Section 4.2).  In particular, the
low-density tail of the initial PDF remains Gaussian down to
fluctuations of $\sim$2.5 times the standard deviation.  The fact that
`the voids are too devoid' of bright galaxies has often been seen as a
possible problem for CDM-like models of structure formation (see,
e.g., Peebles 1999).  The result presented here shows that the voids,
as defined by IRAS galaxies, are as underdense as they are expected to
be as an effect of gravitation, at least on scales as small as 5 \mpc,
where the highly non-linear regime is still not dominant.

\section{Summary and Conclusions}

The ZTRACE algorithm has been applied to the new IRAS PSCz redshift
catalogue to reconstruct the initial conditions of our local Universe.
The ZTRACE algorithm, recently proposed by ME99, is known to give an
unbiased estimate of the initial density which generates, through
gravitational collapse in the Zel'dovich (1970) approximation, the
observed galaxy density field in the redshift space under the
assumption that galaxies trace mass.  The reconstructed initial
conditions of our local Universe can be used as input to N-body
simulations, with the small-scale modes restored with a Monte-Carlo
procedure (Kolatt et al. 1996; Narayanan et al. 1999).

The results presented in this paper push the reconstruction of initial
conditions to scales of 5 \mpc.  This is close to the limit at which
highly non-linear dynamics becomes dominant, hampering any detailed
inversion of the gravitational evolution.  This paper improves by a
factor of two in scale over the previous determination of the initial
PDF, by Nusser et al. (1995).  This improvement is due to the
increased sampling density and depth of the new PSCz catalogue
relative to previous IRAS-based catalogues such as QDOT (Lawrence et
al. 1999) or 1.2 Jy Fisher et al. (1995), and to the superior
performance of ZTRACE, compared to previous reconstruction algorithms
(see ME99 and Narayanan \& Croft 1999).

The reconstructed initial PDF of the local Universe is Gaussian to
within the accuracy of the method.  The lack of detection of
non-Gaussianity in the initial PDF supports the hypothesis that
initial conditions are Gaussian and IRAS galaxies are not strongly
biased with respect to the matter density.  Moreover, it shows that
the voids are as devoid of bright galaxies as they should be as an
effect of gravity (see, e.g., Peebles 1999).  The Gaussian nature of
the primordial perturbations has been tested on much larger scales and
with greater accuracy by measurements of the CMB.  Then this result,
besides offering a complementary support for Gaussianity on the scale
of $\sim$5--10 \mpc, gives original constraints on the nature of
galaxy bias, which in general induces distortions on the initial PDF
reconstructed under the no-bias hypothesis.  However, our tests show
that the distortion induced by realistic power-law galaxy bias models,
especially on the low-density tail, is just beyond the reach of the
PSCz catalogue.  Future surveys, like the 2dF or the SDSS, covering
larger volume than PSCz, will provide a way to constrain galaxy bias
beyond the linear level from the reconstruction of the initial PDF.

\begin{figure*}
\centerline{
\psfig{figure=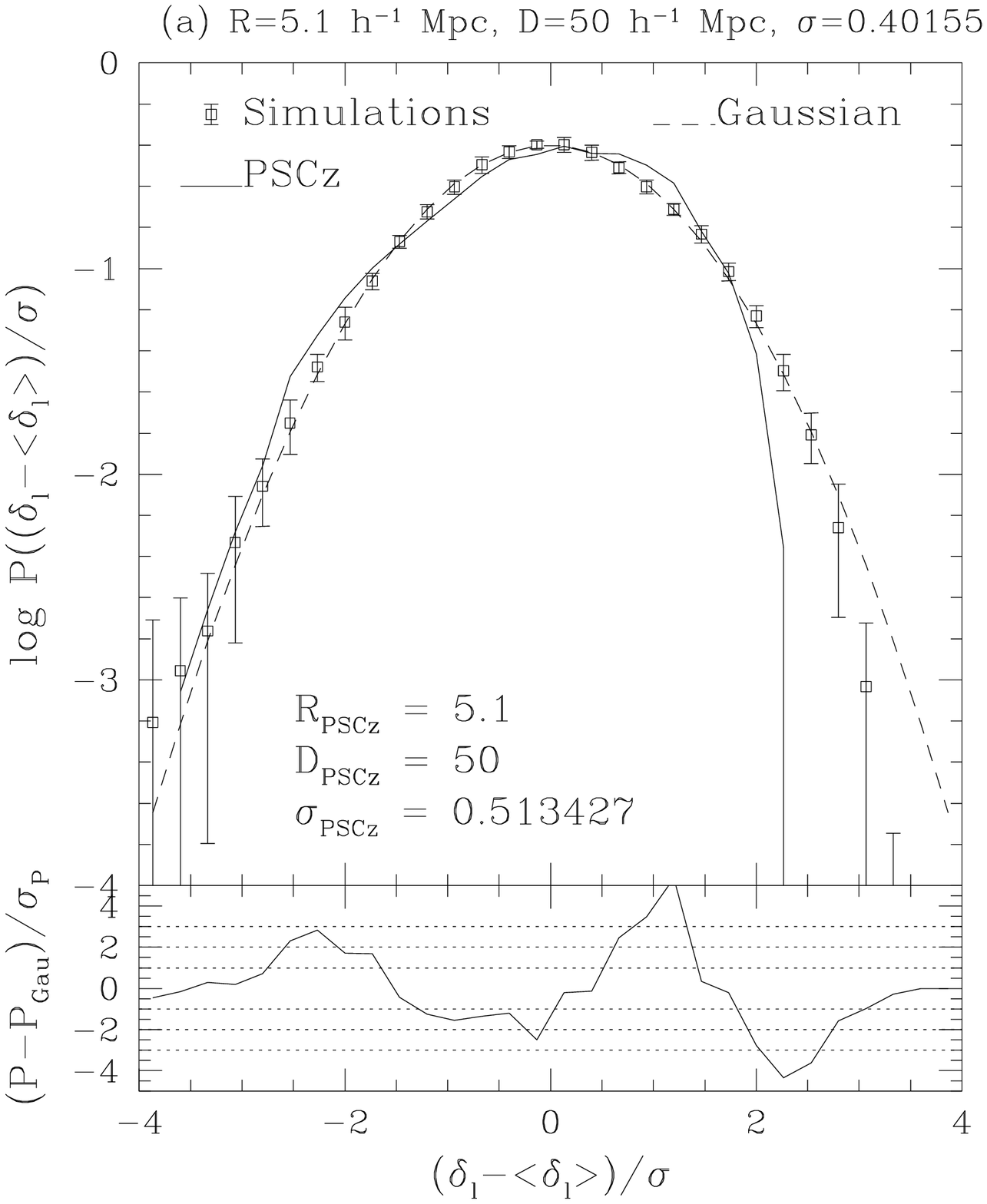,width=10cm}
\psfig{figure=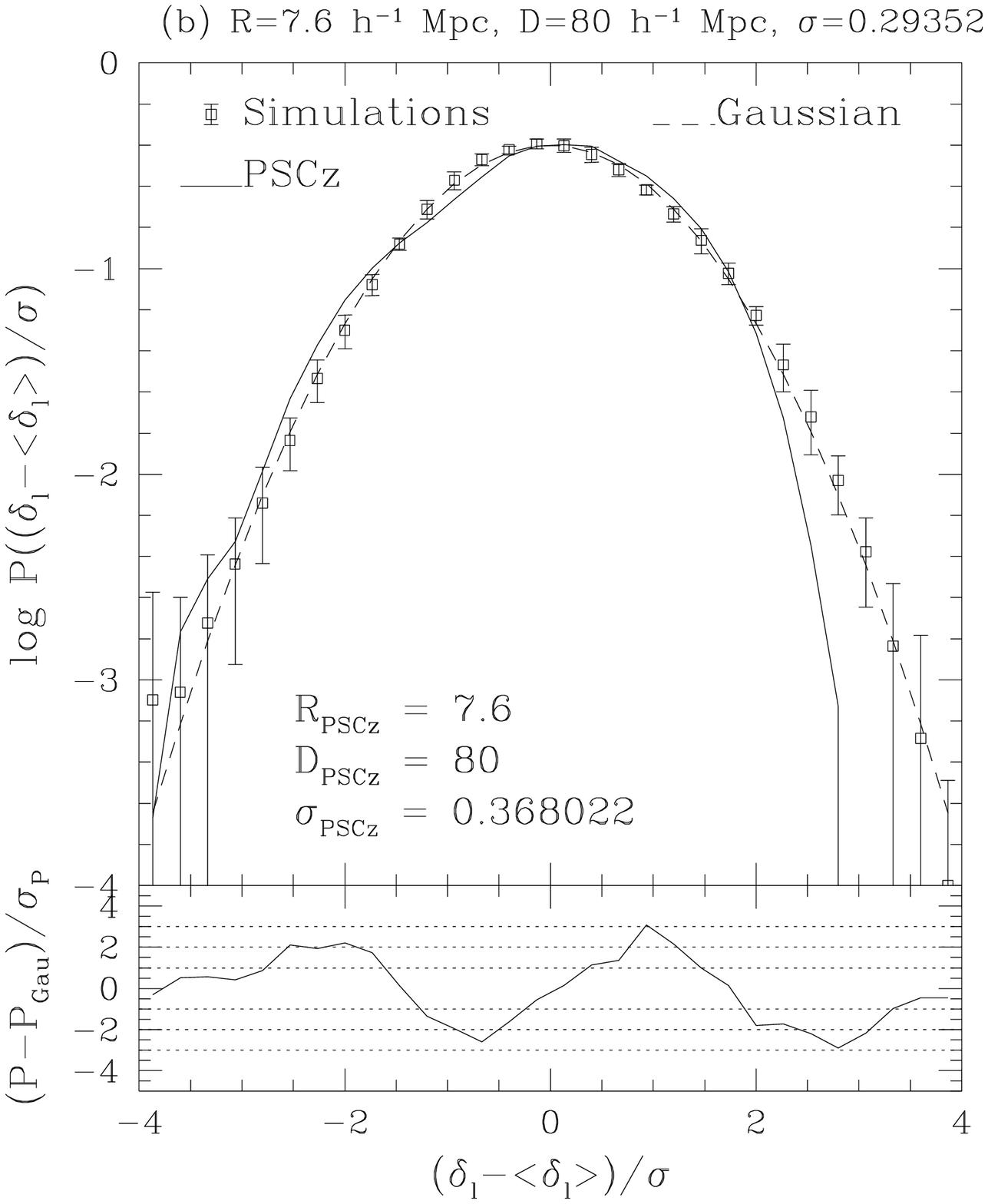,width=10cm}}
\caption{Reconstructed initial PDFs for the PSCz catalogue compared
with those for the simulated catalogues.  The smoothing radii are
$R=5.1$ and 7.6 \mpc\ within $D=50$ and 80 \mpc.  The dashed Gaussian
line is shown for reference.  The lower panels show the significance
of the disagreement.}
\end{figure*}

\begin{figure*}
\centerline{
\psfig{figure=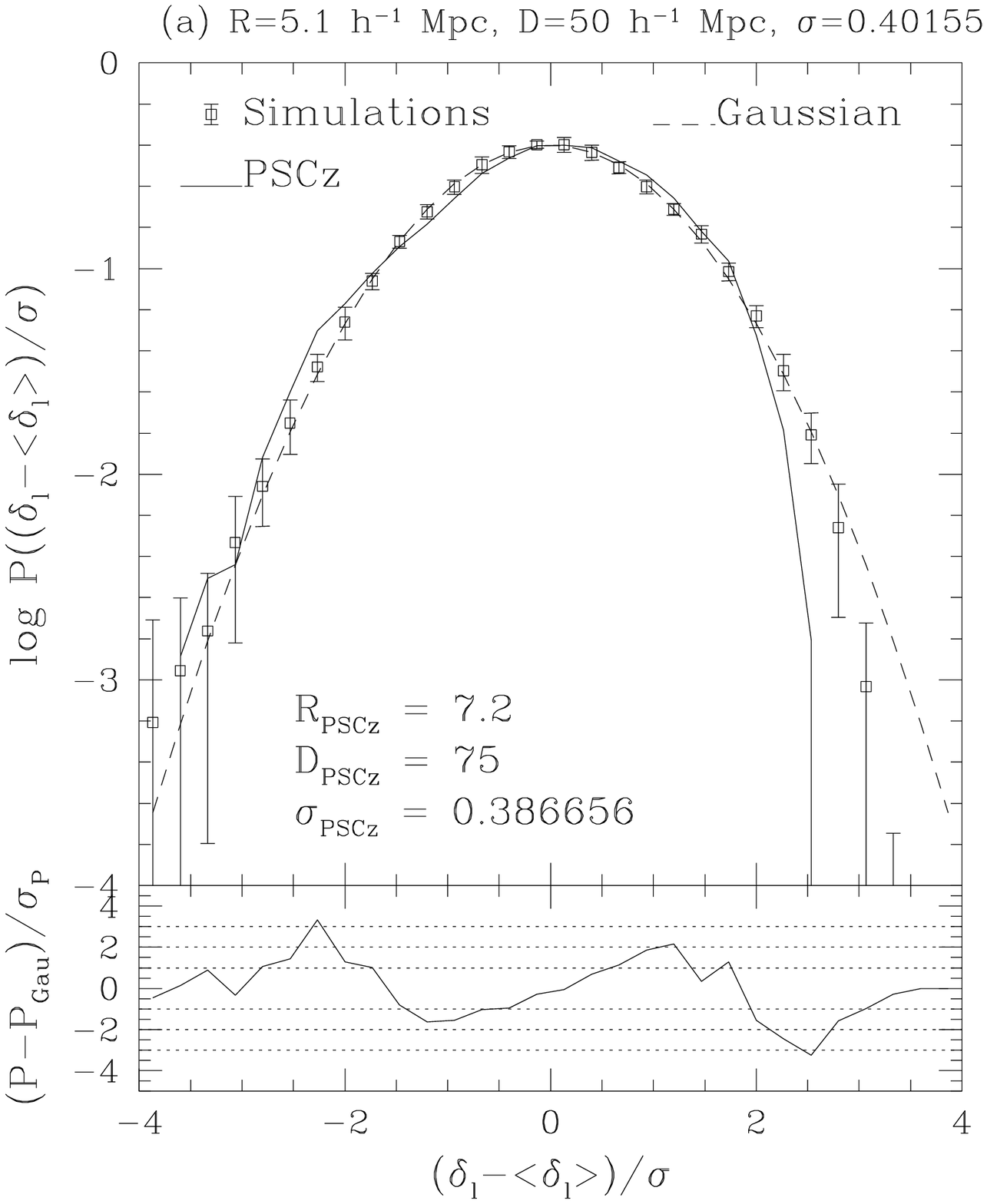,width=10cm}
\psfig{figure=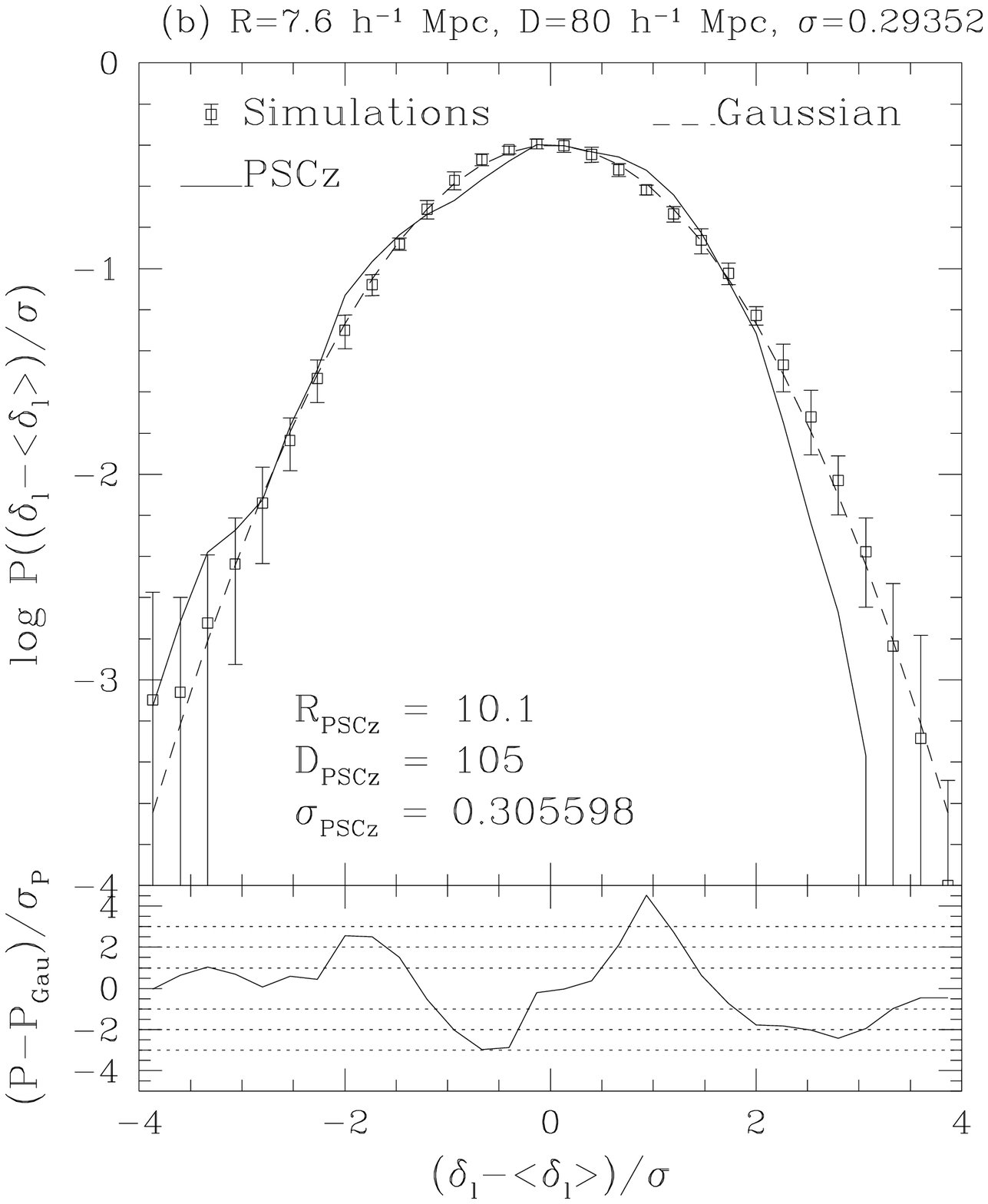,width=10cm}}
\caption{As in Fig. 8, but the PSCz catalogue is smoothed with $R=7.2$
and 10.1 \mpc\ within $D=75$ and 105 \mpc.  The smoothing radii are
chosen so that the variances of the initial PDFs for PSCz and
simulated catalogues roughly coincide.}
\end{figure*}

\section*{Acknowledgments}

The Hydra team (H. Couchman, P. Thomas, F. Pearce) has provided the
code for N-body simulations.  Volker Springel has kindly provided a C
version of the code for adaptive smoothing.  We thank the referee,
David Weinberg, for his helpful comments.  P.M. thanks Christian
Marinoni and Inga Schmoldt for discussions.  P.M.  has been supported
by the EC TMR Marie Curie grant ERB FMB ICT961709.

\bsp
\label{lastpage} 
\end{document}